\documentclass[12pt]{article}

\usepackage{amsfonts}
\usepackage{amsmath}
\usepackage{epsfig}
\usepackage{latexsym}
\usepackage{graphicx}
\usepackage{color}
\usepackage{soul}
\usepackage{cancel}
\usepackage{amssymb}
\usepackage{epstopdf}
\numberwithin{equation}{section}
\allowdisplaybreaks

\def\spa#1{\phantom{\fbox{\rule[-#1cm]{0cm}{0cm}}}}

\def\be{\begin{equation}}
\def\ee{\end{equation}}
\def\bea{\begin{eqnarray}}
\def\eea{\end{eqnarray}}
\def\bequ{\begin{equation}}
\def\eequ{\end{equation}}

\def\half{{1\over 2}}
\def\Tr{\mbox{Tr}}
\def\del{\partial}

\renewcommand{\thefootnote}{\fnsymbol{footnote}}

\newcommand{\eq} {equation}
\newcommand{\eqa} {eqnarray}
\newcommand{\NN} {\nonumber}

\usepackage{latexsym}\usepackage{epsfig,amssymb,euscript}
\usepackage{amsmath, mathrsfs} \usepackage{bbm, slashed}
\usepackage{color}
\usepackage{ulem}
\usepackage{cite}
\usepackage{graphicx}

\newfam\frakfam
\font\teneufm=eufm10
\font\seveneufm=eufm7
\font\fiveeufm=eufm5
\textfont\frakfam=\teneufm
\scriptfont\frakfam=\seveneufm
\scriptscriptfont\frakfam=\fiveeufm

\def\bb{
\font\tenmsb=msbm10
\font\sevenmsb=msbm7
\font\fivemsb=msbm5
\textfont1=\tenmsb
\scriptfont1=\sevenmsb
\scriptscriptfont1=\fivemsb
}

\newfam\mbffam
\font\tenmbf=cmmib10
\font\sevenmbf=cmmib7
\font\fivembf=cmmib5
\textfont\mbffam=\tenmbf
\scriptfont\mbffam=\sevenmbf
\scriptscriptfont\mbffam=\fivembf

\newfam\mbfcalfam
\font\tenmbfcal=cmbsy10
\font\sevenmbfcal=cmbsy7
\font\fivembfcal=cmbsy5
\textfont\mbfcalfam=\tenmbfcal
\scriptfont\mbfcalfam=\sevenmbfcal
\scriptscriptfont\mbfcalfam=\fivembfcal

\newfam\mscrfam
\font\tenmscr=rsfs10
\font\sevenmscr=rsfs7
\font\fivemscr=rsfs5
\textfont\mscrfam=\tenmscr
\scriptfont\mscrfam=\sevenmscr
\scriptscriptfont\mscrfam=\fivemscr

\def\tilde{\widetilde}

\def\bar{\overline}
\def\b{\bar}
\def\bsq#1{{{\b{#1}}^{\lower 2.5pt\hbox{$\scriptstyle 2$}}}}
\def\bexp#1#2{{{\b{#1}}^{\lower 2.5pt\hbox{$\scriptstyle #2$}}}}
\def\dotexp#1#2{{{#1}^{\lower 2.5pt\hbox{$\scriptstyle #2$}}}}

\def\rt2{\sqrt{2}}
\def\half {{1 \over 2}}
\def\Re{\mathop{\rm Re}}

\def\Tr{\mathop{\rm Tr}}

\def\underrel#1\over#2{\mathrel{\mathop{\kern\z@#1}\limits_{#2}}}

\font\tenbifull=cmmib10
\font\tenbimed=cmmib7
\font\tenbismall=cmmib5
\mathchardef\bbGamma="7000
\mathchardef\bbDelta="7001
\mathchardef\bbPhi="7002
\mathchardef\bbAlpha="7003
\mathchardef\bbXi="7004
\mathchardef\bbPi="7005
\mathchardef\bbSigma="7006
\mathchardef\bbUpsilon="7007
\mathchardef\bbTheta="7008
\mathchardef\bbPsi="7009
\mathchardef\bbOmega="700A
\mathchardef\bbalpha="710B
\mathchardef\bbbeta="710C
\mathchardef\bbgamma="710D
\mathchardef\bbdelta="710E
\mathchardef\bbepsilon="710F
\mathchardef\bbzeta="7110
\mathchardef\bbeta="7111
\mathchardef\bbtheta="7112
\mathchardef\bbiota="7113
\mathchardef\bbkappa="7114
\mathchardef\bblambda="7115
\mathchardef\bbmu="7116
\mathchardef\bbnu="7117
\mathchardef\bbxi="7118
\mathchardef\bbpi="7119
\mathchardef\bbrho="711A
\mathchardef\bbsigma="711B
\mathchardef\bbtau="711C
\mathchardef\bbupsilon="711D
\mathchardef\bbphi="711E
\mathchardef\bbchi="711F
\mathchardef\bbpsi="7120
\mathchardef\bbomega="7121
\mathchardef\bbvarepsilon="7122
\mathchardef\bbvartheta="7123
\mathchardef\bbvarpi="7124
\mathchardef\bbvarrho="7125
\mathchardef\bbvarsigma="7126
\mathchardef\bbvarphi="7127

\def\CA{{\cal A}}

\def\CD{{\cal D}}

\def\CL{{\cal L}}
\def\CM{{\cal M}}
\def\CN{{\cal N}}
\def\CO{{\cal O}}

\def\CW{{\cal W}}

\def\1{{\ds 1}}

\def\Z{\hbox{$\bb Z$}}

\begin{document}
\title{Cardy Formula for 4d SUSY Theories and Localization}
\author{
Lorenzo Di Pietro$^{a,b}$\footnote{ldipietroATperimeterinstitute.ca}, $\ $  
Masazumi Honda$^{b}$\footnote{masazumi.hondaATweizmann.ac.il} 
  \spa{0.5} \\
\\
$^{a}${\small{\it Perimeter Institute for Theoretical Physics,}}
\\ {\small{\it Waterloo, Ontario, Canada}} \\
$^{b}${\small{\it Department of Particle Physics and Astrophysics,}}
\\ {\small{\it Weizmann Institute of Science, Rehovot 7610001, Israel}} \\
}
\date{\small{November 2016}}

\maketitle
\thispagestyle{empty}
\centerline{}

%%%%%%%%%%%%%%%%%%%%%%%%%%%%%%%%%%%%%%
\begin{abstract}

We study 4d $\mathcal{N}=1$ supersymmetric theories on a compact Euclidean manifold of the form $S^1 \times\mathcal{M}_3$. Partition functions of gauge theories on this background can be computed using localization, and explicit formulas have been derived for different choices of the compact manifold $\CM_3$. Taking the limit of shrinking $S^1$, we present a general formula for the limit of the localization integrand, derived by simple effective theory considerations, generalizing the result of \cite{Ardehali:2015bla}. The limit is given in terms of an effective potential for the holonomies around the $S^1$, whose minima determine the asymptotic behavior of the partition function. If the potential is minimized in the origin, where it vanishes, the partition function has a Cardy-like behavior fixed by $\Tr(R)$, while a nontrivial minimum gives a shift in the coefficient. In all the examples that we consider, the origin is a  minimum if $\Tr(R) \leq 0$.

\end{abstract}
\vfill
\noindent 
{\small WIS/07/16-OCT-DPPA}

\renewcommand{\thefootnote}{\arabic{footnote}}
\setcounter{footnote}{0}
\newpage
\setcounter{page}{1}
\tableofcontents

%%%%%%%%%%%%%%%%%%%%%%%%%%%%%%%%
%%%%%%%%%%%%%%%%%%%%%%%%%%%%%%%%
%%%%%%%%%%%%%%%%%%%%%%%%%%%%%%%%
\section{Introduction}
%%%%%%%%%%%%%%%%%%%%%%%%%%%%%%%%
%%%%%%%%%%%%%%%%%%%%%%%%%%%%%%%%
%%%%%%%%%%%%%%%%%%%%%%%%%%%%%%%%
An interesting class of observables in supersymmetric quantum field theory is given by the Euclidean partition functions on $S^1 \times \CM_{d-1}$, where $\CM_{d-1}$ is a compact $(d-1)$-dimensional manifold. 
With some restrictions on the manifold $\CM_{d-1}$, theories in $2 \leq d \leq 4$ with four supercharges and a $U(1)_R$ symmetry can be coupled to $S^1 \times \CM_{d-1}$ preserving at least two supercharges of opposite $R$-charge $(Q, \tilde{Q})$. (A classification of the allowed $\CM_{d-1}$ for $d=3,4$ can be found in \cite{Dumitrescu:2012ha,Klare:2012gn,Closset:2012ru}.) The resulting partition functions, up to a Casimir energy factor \cite{Kim:2012ava, Assel:2015nca, Bobev:2015kza, Martelli:2015kuk}, count the states in the Hilbert space of the theory on $\CM_{d-1}$ that are annihilated by $\{ Q , \tilde{Q}\}$, weighted by the fermion number $(-1)^F$. 
From the index interpretation 
it follows that they are independent on continuous coupling constants \cite{Witten:1982df}, 
therefore their computation at weak coupling is valid even when the coupling is large. 
Exact results on several different geometries exist 
in the literature \cite{Kinney:2005ej, Romelsberger:2005eg, Bhattacharya:2008zy, Closset:2013sxa, Benini:2015noa, Honda:2015yha, Benini:2011nc, Razamat:2013opa, Assel:2014paa, Benini:2016hjo, Closset:2016arn, Nishioka:2014zpa}, 
derived either via supersymmetric localization \cite{Pestun:2007rz, Pestun:2016zxk} or 
by solving the associated counting problem in the Hilbert space.

In this work we will focus on $\CN = 1$ theories in $d=4$, and study the behavior of such partition functions in the limit in which the length $\beta$ of the circle goes to zero. Similarly to the Cardy formula in $2d$ CFT \cite{Cardy:1986ie}, this limit controls the asymptotic behavior at large energies of the weighted density of short states. Following the observations in \cite{Aharony:2013dha, Ardehali:2013gra, Ardehali:2013xya}, and assuming the existence of a weakly-coupled point in the space of couplings,\footnote{The derivation in \cite{DiPietro:2014bca} used the existence of a certain mixed Chern-Simons term of order $\frac{1}{\beta}$ in the effective action on $\CM_3$ in the limit $\beta \to 0$. The existence of this term has been recently related to global anomalies in \cite{Golkar:2015oxw}. This allows to fix the coefficient ${\rm mod}~2$ without assuming the existence of a weakly-coupled point. For other non-perturbative arguments, see \cite{Jensen:2012kj, Jensen:2013kka, Jensen:2013rga}. Evidence for the validity of the Cardy-like behavior \eqref{eq:Cardy} in non-Lagrangian theories was provided in \cite{Buican:2015ina}.} it was argued in \cite{DiPietro:2014bca}  that for $\beta \to 0$ the partition function has a universal behavior (see also \cite{Ardehali:2014zba, Ardehali:2014esa, Ardehali:2015hya, Shaghoulian:2015kta, Shaghoulian:2015lcn})
\be
Z_{S^1 \times \CM_3} \underset{\beta \to 0}{\longrightarrow} e^{-\frac{\pi^2\Tr(R)L_{\CM_3}}{12 \beta} +\,\dots} \times Z_{\CM_3}~. \label{eq:Cardy}
\ee 
$\Tr(R)$ is the anomaly coefficient of the $U(1)_R$ symmetry used to couple the theory to the curved background, $L_{\CM_3}$ is the integral on $\CM_3$ of a local functional of the background fields in the $3d$ supergravity multiplet\footnote{We are choosing a different normalization for $L_{\CM_3}$ compared to \cite{DiPietro:2014bca}, namely $L_{\CM_3}^{\rm here} = 12 L_{\CM_3}^{\rm there}$~.}, $Z_{\CM_3}$ denotes the partition function on $\CM_3$ of the dimensional reduction of the theory, and the ellipses stand for subleading $\beta$-dependent terms. 

In a weakly-coupled point, using localization, $Z_{S^1\times \CM_3}$ and $Z_{\CM_3}$ can be written as an integral over the maximal torus of the gauge group and the Cartan subalgebra of the corresponding Lie algebra, respectively.  If we use these formulas to evaluate \eqref{eq:Cardy} on $\CM_3 = S^3$ we may run into a problem:  in general the integral expression for $Z_{S_3}$ converges only for $R$-charge assignments in a certain range\cite{Willett:2011gp,Morita:2011cs,Safdi:2012re,Lee:2016zud}. In the typical examples of gauge theories with vector-like matter in the fundamental representation, the $4d$ non-anomalous $R$-symmetry falls in the range that makes the $Z_{S_3}$ integral convergent, but in some more exotic examples \cite{Intriligator:1994rx, Brodie:1998vv} it falls outside and the resulting integral is exponentially divergent. The possible divergences manifest in the limit $\beta \to 0$ through a modification of the asymptotic formula \eqref{eq:Cardy}. This was first observed in\cite{Ardehali:2015bla}.

In this paper we will study this phenomenon in the context of the general $S^1 \times \CM_3$ partition function. We will use that in the limit $\beta \to 0$ the integrand of the localization formula for $Z_{S^1 \times \CM_3}$ reduces to a simple ``effective potential'' for the holonomies 
\begin{align}
Z_{S^1 \times \CM_3} & \underset{\beta \to 0}{\longrightarrow}e^{ -\frac{\pi^2\Tr(R)L_{\CM_3}}{12 \beta}} \int d^{r} a~e^{- V^{\text{eff}}_{\CM_3}(a) +\,\dots}~.\label{eq:limitint}
\end{align}
Here $e^{2\pi i a_i}$ are the holonomies around the $S^1$, valued in the maximal torus of the gauge group $G$, with $i=1,\dots,r={\text{rank}(G)}$. $V^{\text{eff}}_{S^3}(a)$ was computed in \cite{Ardehali:2015bla} starting from the integral representation of the partition function on $S^1\times S^3$. We will show an alternative way to compute the effective potential, that does not rely on the knowledge of the matrix model, and readily generalizes to an arbitrary supersymmetry-preserving $\CM_3$. Our computation relies on the $3d$ effective-theory approach that was used to derive \eqref{eq:Cardy}. The resulting formula is 
\begin{align}
V^{\text{eff}}_{\CM_3}(a)   
=  & - \sum_f \sum_{\rho_f \in \mathfrak{R}_f} \left[\frac{\pi^3 i A_{\CM_3}}{6 \beta^2}\kappa(\rho_f \cdot a)  + \frac{\pi^2 ( R_f L_{\CM_3} - \rho_f \cdot l_{\CM_3})}{2 \beta}\vartheta(\rho_f \cdot a) \right]~,\nonumber\\
& \kappa(x)  \equiv \{x\}(1-\{x\})(1-2\{x\})~, \quad
\vartheta(x)  \equiv \{x\}(1-\{x\})~.\label{eq:effpot}
\end{align}
Here $\{x\}$ denotes the fractional part of $x$. The sum runs over all the fermions $f$ with $R$-charge $R_f$, $\mathfrak{R}_f$ being their representation under the gauge group and $\rho_f$ the associated weights. $A_{\CM_3}$, $L_{\CM_3}$ and $l^i_{\CM_3}$, $i=1,\cdots,r$ are given by integrals on $\CM_3$ of certain local densities that we will specify. 
We compared this expression with the available localization formulas, and in all cases we found agreement.  For the cases in which explicit localization formulas have not been derived yet, our result gives a constraint on the form of the integrand.

Our method to obtain $V^{\text{eff}}_{\CM_3}(a)$ can be summarized as follows: we first take $G$ to be a global symmetry, we turn on fugacities $m_i$ in the Cartan of $G$, and take the limit $\beta \to 0$ together with $m_i\to \infty$ keeping $a_i = \beta m_i$ finite. Each matter multiplet gives rise to a tower of Kaluza-Klein (KK) modes that have large masses in this limit and can be integrated out, generating supersymmetric Chern-Simons terms for the gauge fields in the background supergravity and vector multiplets. The potential is obtained by summing the coefficients of these Chern-Simons terms over the full KK tower, with an appropriate regularization, and evaluating the resulting functionals on the background. In this sense, we can view $V^{\text{eff}}_{\CM_3}(a)$ as a refinement of the Cardy formula of \cite{DiPietro:2014bca} which includes dependence on the fugacities $a_i = \beta m_i$. 
Then, if $G$ is anomaly-free, one can gauge it by coupling it to dynamical gauge fields. At the level of the partition function, this requires to integrate over the $a_i$, and possibly introduce a discrete sum over topological sectors for the gauge fields. Moreover $V^{\text{eff}}_{\CM_3}(a)$ will receive an additional contribution from the KK modes of the vector multiplets, that we can compute in the same way.

Large gauge transformations imply the identification $a_i \sim a_i + 1$. However it is not a priori clear that $V^{\text{eff}}_{\CM_3}(a)$ obtained with the method just described will be a periodic function, because the real masses of the $3d$ KK modes are not periodic functions of $a_i$. Nevertheless we will see that when we regulate the sum over the infinite tower and we evaluate for supersymmetric configurations of the vector multiplets, the dependence on the integer part of $a_i$ drops, making $V^{\text{eff}}_{\CM_3}(a)$ a periodic function. We regard this as a consistency check that our regularization procedure is compatible with supersymmetry and gauge invariance.

In general it is not easy to study the $\beta \to 0$ limit of the integral in \eqref{eq:limitint}, because $V^{\rm eff}_{\CM_3}(a)$ is not a smooth function. If the matter content is symmetric under $\rho_f \leftrightarrow -\rho_f$, the potential reduces to the single term proportional to the density $L_{\CM_3}$. In this case, the analysis of \cite{Ardehali:2015bla} shows that the limit $\beta \to 0$ is dominated by the minima of $V^{\text{eff}}_{\CM_3}(a)$. If $V^{\text{eff}}_{\CM_3}(a)$ has a local minimum in the origin, where it vanishes, then the only contribution in \eqref{eq:limitint} comes from the prefactor, and \eqref{eq:Cardy} is valid. Alternatively, $V^{\text{eff}}_{\CM_3}(a)$ can have a local minimum at some $a_{\text{min}}$ where $V^{\text{eff}}_{\CM_3}(a_{\text{min}}) < 0$. In this case \eqref{eq:Cardy} is amended by an additional term that goes like $e^{-V^{\text{eff}}_{\CM_3}(a_{\text{min}})}$ for $\beta \to 0$. The intermediate case in which $V^{\text{eff}}_{\CM_3}(a)$ is flat leads to additional powers of $\beta$ in \eqref{eq:Cardy}. In \cite{Ardehali:2015bla} it was also shown that when $V^{\text{eff}}_{S^3}(a)$ has a local minimum in the origin then the integrand of $Z_{S^3}$ is damped exponentially at infinity, while when $V^{\text{eff}}_{S^3}(a_{\text{min}})<0$ it grows exponentially. In this paper we will generalize this result to the case where $\CM_3$ is a Lens space or $S^1\times \Sigma_g$, $\Sigma_g$ being a Riemann surface of genus $g$.

There are few known examples \cite{Intriligator:1994rx, Brodie:1998vv} so far of theories with $V^{\rm eff}_{\CM_3}(a_{\rm min}) < 0$, and they share some intriguing features: 
1) the unbroken $R$-symmetry has $\Tr(R) > 0$, which implies that in absence of emergent symmetries the IR SCFTs will have $c - a<0$ ; 
2) they are examples of so-called ``misleading anomaly matching'' \cite{Intriligator:1994rx}, 
meaning that all 't~Hooft anomalies are matched by a putative confined phase, 
but various arguments rule out this possibility and 
point to the existence of an interacting IR phase (see also \cite{Intriligator:2005if, Poppitz:2009kz,Vartanov:2010xj, Gerchkovitz:2013zra}). We will analyze a large class of examples and find that in all these theories the potential is minimized in the origin %if and only if %
when $\Tr(R) \leq 0$.\footnote{In the first version we claimed that all the known examples satisfy that the potential is minimized in the origin if and only if $\Tr(R) \leq 0$. Later we became aware of a counterexample to this stronger statement in the context of class S theories, discussed in \cite{Ardehali:2016kza}. We thank A. A. Ardehali for a discussion on this point. \label{foot:cexample}} We will also find new examples of theories in the class with nontrivial minimum. 

The rest of the paper is organized as follows: section \ref{sec:eff} contains the derivation of the potential for the holonomies using the effective theory approach, section \ref{sec:loc} presents the comparison with the available localization results, in section \ref{sec:min} we derive the relation between the minima of the effective potential and the behavior of the integrand in $Z_{\CM_3}$, section \ref{sec:sign} contains the examples which demonstrate the connection to the sign of $\Tr(R)$, and finally section \ref{sec:disc} contains a summary and a discussion of possible future directions.
 
%%%%%%%%%%%%%%%%%%%%%%%%%%%%%%%%%%% 
\section{Effective Potential for the Holonomies}\label{sec:eff}
%%%%%%%%%%%%%%%%%%%%%%%%%%%%%%%%%%%%
In this section we derive the formula \eqref{eq:effpot} for the effective potential for the holonomies in a generic supersymmetric background $S^1\times \CM_3$. We will start by considering the Chern-Simons effective action induced by the dimensional reduction of fermions coupled to background gauge fields. We will then supersymmetrize the effective action, and show that this leads to \eqref{eq:effpot}. Finally we will evaluate it in the examples $\CM_3 = S^3_b~, L(n,1),~S^1\times \Sigma_g$. 

%%%%%%%%%%%%%%%%%%%%%%%%%%%%%
\subsection{Chern-Simons Terms from KK Fermions}\label{sec:CSKK}
%%%%%%%%%%%%%%%%%%%%%%%%%%%%%
We consider $4d$ Weyl fermions on $S^1 \times \mathbb{R}^3$ minimally coupled to background gauge fields $A = A_M dx^M$, $M = 1,2,3,4$ for a symmetry group $G$. We denote with $f$ an index running over the fermions $\Psi_f$, each of which is in a  representation $\mathfrak{R}_f$ of $G$. 

We turn on holonomies around the $S^1$, namely $\langle A \rangle = (A_4^1,\dots, A^r_4) dx^4$ where $r$ is the rank of $G$.
$x^4$ is the circle coordinate, subject to the identification $x^4 \sim x^4 + \beta$. The holonomies break $G$ to the maximal torus $U(1)^r$. We also turn on a background metric
\be
g_{MN}dx^M dx^N = (dx^4 + c_\mu dx^\mu)^2 + h_{\mu\nu}dx^\mu dx^\nu~,~~\mu=1,2,3~.
\label{eq:bckgmetr}
\ee
$c_\mu$ is a $3d$ gauge field for the $U(1)_{KK}$ isometry associated to the circle. The combination
\be
\CA =(A_\mu - c_\mu A_4) dx^\mu~,
\ee 
is invariant under $U(1)_{KK}$ and can be thought of as a $3d$ gauge field.

Starting from the free $4d$ Lagrangian 
\be
i \bar \Psi_f \bar{\sigma}^M (\nabla_M  \Psi)_f~,
\ee 
where $\nabla_M$ contains minimal coupling to the the metric and the $G$ gauge fields, we get an infinite tower of $3d$ KK fermions $\psi_f^n$ coupled to the $3d$ metric and $G\times U(1)_{KK}$ gauge fields. 
Due to the $G$-holonomies, a KK mode $\psi^n_f$ in a certain representation $\mathfrak{R}_f$ is split into components $\{\psi^n_{\rho_f}\}_{\rho_f \in \mathfrak{R}_f}$ of definite real mass $M_{\rho_f}^n = {2 \pi n\over \beta} + \rho_f \cdot A_4$ and $U(1)^r$ charges $\rho_f$, each associated to one of the weights $\rho_f$ of the representation $\mathfrak{R}_f$. Here $\cdot$ denotes the inner product in the Cartan subalgebra of the Lie algebra of $G$. The number $n$ is integer or half-integer depending on the periodicity conditions, and it also corresponds to the charge of $\psi^n_{\rho_f}$ under $U(1)_{KK}$.

Consider the generating functional  $\CW \equiv - \log Z$. Following \cite{DiPietro:2014bca, Golkar:2012kb} (see also \cite{Grimm:2011fx, Cvetic:2012xn, Grimm:2015zea}), we can determine its dependence on the $U(1)^r \times U(1)_{KK}$ gauge fields by integrating out the tower of massive KK fermions. The $\beta$-dependent part of $\CW$ can be expanded in derivatives, and the leading contribution for $\beta \to 0$ comes from the Chern-Simons terms. As shown in \cite{DiPietro:2014bca, Golkar:2012kb}, the sum over KK modes leads to
\begin{align}
& \CW \supset -{i\over 8 \pi} \sum_f \sum_{\rho_f \in \mathfrak{R}_f} \int\nonumber\\& \left[ S_1(\rho_f \cdot a) (\rho_f \cdot \CA) \wedge d (\rho_f \cdot \CA)  - 2 S_2(\rho_f \cdot a) {2 \pi \over \beta}  (\rho_f \cdot \CA)\wedge d c  +S_3(\rho_f \cdot a) {4 \pi^2 \over \beta^2} c\wedge d c \right] ~. \label{eq:overn}
\end{align}
The functions $S_{1,2,3}(x)$ are obtained by continuing to $s=0$ the infinite sums
\be
S_k(s, x)  = \sum_{n \in \Z /   \Z + \half } {\rm sgn}(n + x) n^{k-1}|n + x|^{-s}~,~~k=1,2,3~,\label{eq:sums}  
\ee
for which a closed formula can be written in terms of Hurwitz $\zeta$-functions. Note that decomposing $x = [x] + \{x\}$,
where the square/curly brackets denote the integer/fractional part, we have
\begin{align}
S_1(s, x) & = S_1(s, \{x\})~, \label{eq:intpartS1}\\
S_2(s, x) & = S_2(s, \{x\}) - [x] S_1(s, \{x\})~, \label{eq:intpartS2}\\
S_3(s, x) & = S_3(s, \{x\}) - 2 [x] S_2(s, \{x\}) + [x]^2 S_1(s, \{x\})~. \label{eq:intpartS3}
\end{align}
The result of the $\zeta$-function regularization for periodic conditions is
\begin{align}
S_1(\{x\}) & =  1- 2\{x\} ~,  \label{eq:resp1}  \\
S_2(\{x\}) & = - \frac 16 + \{x\}^2 ~, \label{eq:resp2}\\
S_3(\{x\}) & = - \frac 23 \{x\}^3~. \label{eq:resp3}
\end{align}
For future reference, from eq.s \eqref{eq:intpartS1}, \eqref{eq:intpartS2} and \eqref{eq:intpartS3} we note that there are two linear combinations of these functions which are periodic in $x$ with period 1, namely
\begin{align}
\kappa(x) & \equiv -  3 (S_3(x) +   2 x S_2(x)  + x^2 S_1(x))  = \{x\}(1-\{x\})(1-2\{x\})~,\label{eq:kappa}\\
\vartheta(x) & \equiv S_2(x) +    x S_1(x) + \tfrac 16 = \{x\}(1-\{x\}) ~. \label{eq:vartheta}
\end{align}
These functions have definite parity properties
\be
\vartheta(-x) = \vartheta(x)~,~~\kappa(-x) = -\kappa(x)~.
\ee
The Chern-Simons action \eqref{eq:overn} determines the leading dependence of the free fermion generating functional on the background gauge fields $\CA_\mu$ in the limit $\beta \to 0$, with the holonomy $a$ kept fixed. However in general $\mathcal{W}$ will have additional divergent terms in the limit $\beta\to 0$, which can depend on $a$ and on the background fields $h_{\mu\nu}$, $\CA_\mu$ and $c_\mu$ (this is why we used the symbol $\supset$). Using the fact that $\CW$ is a local functional in this limit, and dimensional analysis, one can easily write down all possible divergent terms that can be generated integrating out the KK modes. These additional terms are
\be
\int d^3x \sqrt{h} \left[\left(\frac{2\pi}{\beta}\right)^3 P(a) + \left(\frac{2\pi}{\beta}\right) \left( c_1(a) R + c_2(a) v_\mu v^\mu \right)\right] + \mathcal{O}(\log \beta)~,\label{eq:addterm}
\ee 
where $P(a)$, $c_1(a)$ and $c_2(a)$ are functions of $a$, $R$ is the Ricci scalar of $h_{\mu\nu}$, and $v_\mu = -i \epsilon_{\mu\nu\rho}\partial^\nu c^\rho$ is the dual field strength of $c_\mu$. (Recall that $c_\mu$ is dimensionless, being a component of the metric.)

As observed in \cite{DiPietro:2014bca} (and see references therein), even though \eqref{eq:overn} has been derived in a free theory, the result is valid for a generic Lagrangian theory, even at strong coupling. This is because the Chern-Simons term cannot depend on continuous coupling constants, so we can always tune to the free point, where only the free fermions contribute, and the result is valid in general. However this sort of ``non-renormalization theorem" is only valid for the Chern-Simons term, and not for the additional terms \eqref{eq:addterm}. Therefore, even in the limit  $\beta \to 0$ we cannot obtain the full dependence of $\mathcal{W}$ on $a$. As we will show in the next section, the situation improves if we invoke supersymmetry, because the coefficients of different terms are related to each other by supersymmetry.
   
%%%%%%%%%%%%%%%%%%%%%%%%%
\subsection{Supersymmetrized Effective Action}
%%%%%%%%%%%%%%%%%%%%%%%%%

We will now apply the result of the previous section in the context of an $\CN = 1$ supersymmetric theory with a $U(1)_R$ $R$-symmetry on the manifold $S^1 \times \CM_3$. 

Let us sketch the logic that we follow here: consider first a theory with only chiral multiplets and a flavor symmetry group $G$, and study the supersymmetric partition function as a function of background holonomies $a = \frac{\beta A_4}{2 \pi}$ in the Cartan of $G$. The result of the previous section implies that the partition function contains a singular contribution in the limit $\beta \to 0$ with $a$ fixed, given by the supersymmetrization of \eqref{eq:overn} evaluated on the supersymmetric background. We will argue that, thanks to supersymmetry, this exhausts all singular terms in the limit. Assuming anomaly cancellation, we can then gauge $G$ by coupling it to dynamical vector multiplets. Via localization, the partition function is now given by an integral over $a$. Using again the results of the previous section we can simply add the contribution of the KK modes of the vector multiplet, and fully determine the terms in the integrand which are singular in the limit $\beta \to 0$.

%%%%%%%%%%%%%%%%%%%%%%%%%%%%%%% 
\subsubsection{Supersymmetric background} 
%%%%%%%%%%%%%%%%%%%%%%%%%%%%%
 
In \cite{Dumitrescu:2012ha} it was shown that in order to preserve supersymmetry the four-manifold $S^1 \times \CM_3$ must be complex. If we require to have two Killing spinors of opposite $R$-charge $(\zeta, \tilde{\zeta})$, it must also admit a holomorphic Killing vector. In this case, which is the one we consider here, the manifold is a torus fibration over a Riemann surface. A supersymmetric theory with an unbroken $U(1)_R$ symmetry can be coupled to this geometry by turning on background values for the bosonic fields in the new minimal supergravity multiplet $(g_{MN}, A^{(R)}_M, C_{MN})$ \cite{Festuccia:2011ws}. Here $C_{MN}$ is a two-form gauge field with associated field strength $V_M = - i \epsilon_{MNRS}\partial^N C^{RS} $. The metric $g_{MN}$ can be taken hermitean. Once the complex structure $J_M^{~~N}$ and the hermitean metric are given, the $R$-symmetry gauge field $A^{(R)}_M$ is fixed,  while $V_M$ depends also on an additional complex scalar function that is annihilated by the holomorphic Killing vector\cite{Dumitrescu:2012ha}. The general supersymmetric configuration for a vector multiplet $(A_M, D)$ is given by
\be
D = - \half J^{MN} F_{MN}~, 
\label{eq:Dflux}
\ee
and in addition we can turn on a flat connection $A^{\text{flat}}_M$. $F_{MN}$ can be non-zero only if the four-manifold has nontrivial two-cycles. The presence of two supercharges $(\zeta, \tilde{\zeta})$ further restricts $F_{MN}$ to have only components on the Riemann surface.

We pick coordinates on $S^1\times \CM_3$ and we parametrize the metric as in eq. \eqref{eq:bckgmetr}. In the limit $\beta \to 0$ it is convenient to reorganize the background fields in $3d$ multiplets. The precise relation between the $4d$ multiplets and the $3d$ ones can be found in appendix D of \cite{Closset:2012ru}. Let us summarize the result
\begin{itemize} 
\item{$(h_{\mu\nu}, \CA^{(R)}_\mu, H, c_\mu)$ with $\CA^{(R)}_\mu \equiv A^{(R)}_\mu + V_\mu$ and $H \equiv A^{(R)}_4 = V_4$, form the bosonic content of the $3d$ new-minimal supergravity multiplet; $v_\mu = -i \epsilon_{\mu\nu\rho} \partial^\nu c^\rho \equiv 2 V_\mu$ is the dual field strength of $c_\mu$.} 
\item{$(\CA_\mu, \sigma, \CD)$ with $\sigma\equiv A_4$, $\CA_\mu \equiv A_\mu - A_4 c_\mu$ and $\CD \equiv D - A_4 H$ form the bosonic content of a $3d$ vector multiplet.}
\end{itemize}
Given a supersymmetric configuration for the $4d$ vector multiplet $(A_M, D)$ on $S^1 \times \CM_3$, we take the limit $\beta \to 0$ with $ A_4  = \frac{2 \pi a }{\beta} \to \infty$ and $a$ fixed, while the other components $A_\mu$ and $D$ remain finite. Equivalently, we can parametrize the supersymmetric configuration for the $3d$ vector multiplet as
\be\label{eq:susy3dvec}
\sigma = \frac{2\pi a}{\beta}~,~~\CA_\mu = A_\mu -  \frac{2\pi a}{\beta} c_\mu~,~~\CD = D -  \frac{2\pi a}{\beta} H~.
\ee
Note that also $\CA_\mu$ and $\CD$ become large in the limit.

%%%%%%%%%%%%%%%%%%%%%%%%%%%%
\subsubsection{Partition function for chiral multiplets with flavor symmetry $G$}
%%%%%%%%%%%%%%%%%%%%%%%%%%%%%%%
For a theory of chiral multiplets with flavor symmetry $G$, we know from the previous section that $\CW = - \log Z^{S^1 \times \CM_3}_{\rm{chi}}$ must contain the Chern-Simons terms \eqref{eq:overn}. We can easily include also the $R$-symmetry gauge-field, by replacing $\rho_f \cdot \CA \to \rho_f \cdot \CA -R_f \CA^{(R)}$, where $R_f$ denotes the $R$-charge of the fermion in consideration. In addition, now $\CW$ must be a supersymmetric local functional. Each one of the resulting Chern-Simons terms can be completed to an independent supersymmetric action. 

In appendix \ref{app:CSactions} we list these actions and we evaluate them on the configuration \eqref{eq:susy3dvec}. The result is that all the Chern-Simons actions, when evaluated on \eqref{eq:susy3dvec}, become linear combinations of the following three actions
\begin{align}
A_{\CM_3} & = \frac{i}{\pi^2}\int_{\CM_3}d^3 x \sqrt{h}  \left[-c^\mu v_\mu+ 2 H \right]~, \label{eq:AM3}\\
L_{\CM_3} & = \frac{1}{ \pi^2}\int_{\CM_3}d^3 x \sqrt{h}  \left[ -\CA^{(R)\mu} v_\mu  +  v^\mu v_\mu - \half H^2  +  \frac 14 R \right]~,\label{eq:LM3}\\ 
l^{i}_{\CM_3} & = \frac{1}{\pi^2}\int_{\CM_3}d^3 x \sqrt{h} \left[-(A^\mu)^i v_\mu +  D^i\right]~.\label{eq:lM3}
\end{align}
$A_{\CM_3}$ carries dimension of an area, while $L_{\CM_3}$ and $l^i_{\CM_3}$ of a length, and $i=1,\dots,r$ is an index in the Cartan of $G$. (We included a factor of $i$ in the definition of $A_{\CM_3}$ in order to make it real on $S^1 \times S^3$, where $H$ is purely imaginary.)

Looking at the supersymmetric Chern-Simons actions in appendix \ref{app:CSactions} we see that the terms of order $\mathcal{O}(\beta^{-1})$ in \eqref{eq:addterm} are included in these actions, and so are also the terms $(2\pi/\beta)^2\,H$ and $(2\pi/\beta) H^2$  constructed out of the  background field $H$, which was not considered in the previous section. Moreover, the ``cosmological constant'' term of order $(2 \pi / \beta)^3$ in \eqref{eq:addterm} cannot be completed to a supersymmetric action, and therefore it must cancel.\footnote{Terms of the form $\beta^{n-3}\sigma^n$ in the $3d$ effective Lagrangian would contribute to $\sim \frac{1}{\beta^3} P(a)$ when evaluated on the background \eqref{eq:susy3dvec}. To understand why these terms do not appear in the supersymmetric Lagrangian, note that $\sigma$ is the bottom component of a real multiplet $\Sigma$. Integrating out KK modes one can generate $D$-term supersymmetric Lagrangians of the form $\beta^{n-2}\CL_D[\Sigma^n]$ which on the supersymmetric background evaluate to $\sim \frac{1}{\beta^2} a^n H + \dots$ . The latter term is included in the KK-KK supersymmetric Chern-Simons action, and therefore the dependence on $a$ of its coefficient is completely fixed by the results of the previous section.} We conclude that all the singular contributions to the $\beta \to 0$ limit of the partition function come from the supersymmetrization of the Chern-Simons terms.

Summing the contributions of the various terms with the coefficients given in \eqref{eq:overn}, we obtain\footnote{
Note that
when we gauge $G$,
we have $\sum_f \sum_{\rho_f \in \mathfrak{R}_f} \rho_f \cdot l_{\CM_3}=0$ 
because of $G$-gravitational$^2$ anomaly cancellation. Therefore in this case the $a$-independent prefactor contains only the term proportional to $\Tr(R)$.
} 
\be
Z^{S^1 \times \CM_3}_{\rm chi}  \underset{\beta\to0}{\longrightarrow}
e^{-\frac{\pi^2 }{12 \beta}\left( \Tr(R)L_{\CM_3} -\sum_f \sum_{\rho_f \in \mathfrak{R}_f} \rho_f \cdot l_{\CM_3} \right)}
 e^{- V^{\rm eff}_{\CM_3}(a)+~\dots}~, \label{eq:limZchi}
\ee
with
\begin{align}\label{eq:result}
 V^{\text{eff}}_{\CM_3}(a)  
=  - \sum_f \sum_{\rho_f \in \mathfrak{R}_f}& \left[\frac{i \pi^3  A_{\CM_3}}{6 \beta^2}\kappa(\rho_f \cdot a) + \frac{\pi^2 ( R_f L_{\CM_3} - \rho_f \cdot l_{\CM_3})}{2 \beta}\vartheta(\rho_f \cdot a) \right]~.
\end{align}
Notably, the integer part of $\rho_f \cdot a$ canceled in the final result, which therefore can be expressed in terms of $\kappa$ and $\vartheta$ in eq.s \eqref{eq:kappa} and \eqref{eq:vartheta}. 
Hence, $V^{\text{eff}}_{\CM_3}(a)$ is periodic in the $a_i$s with period 1, as required by invariance under large gauge transformations. 
Note also that the terms $\kappa(\rho_f \cdot a)$ and $\rho_f^i \vartheta(\rho_f \cdot a)$ are odd under $\rho_f \to - \rho_f$ and therefore vanish in a theory with charge conjugation invariance.

%%%%%%%%%%%%%%%%%%%%%%%%%%%%%%%%%%%%%%%
\subsubsection{Gauging $G$ and localization matrix model}
%%%%%%%%%%%%%%%%%%%%%%%%%%%%%%%%%%%%%%%
We can couple $G$ to dynamical gauge fields if the $G^3$ and $G$-gravitational$^2$ anomalies cancel. 
This requires
\be
\sum_f \sum_{\rho_f \in \mathfrak{R}_f} \rho^i_f\rho^j_f\rho^k_f = 0~,~~\sum_f \sum_{\rho_f \in \mathfrak{R}_f} \rho^i_f=0~,~~i,j,k=1,\dots,r~. \label{eq:anoodd}
\ee
Moreover we must ensure that the $R$-symmetry that we use to couple the theory to the curved background is unbroken by the gauging, i.e. the $U(1)_R-G^2$ anomaly must cancel
\be
\sum_f \sum_{\rho_f \in \mathfrak{R}_f} \rho_f^i \rho_f^j R_f = 0~. \label{eq:anoeven}
\ee
Note that the sum over fermions here includes both the fermions in chiral matter multiplets and the gauginos, for which $\rho_f$ is a root vector of $G$ and $R_f = 1$. Since the sums in \eqref{eq:anoodd} are odd, the contribution of gauginos need not be included, because there is a cancellation between positive and negative roots.

To compute the $S^1 \times \CM_3$ partition function of the theory obtained by gauging $G$ 
we need to path integrate over the vector multiplets. 
Using the $Q$-exact Yang-Mills action to perform localization, 
the path integral is reduced to a sum over the constant supersymmetric configurations 
for the bosonic fields in the vector multiplet. 
Those include constant holonomies around $S^1$, parametrized by the variables $a^i$, $i=1,\dots , r$. If $\pi^1(\CM_3)$ is nontrivial 
we also have to sum over additional holonomies, that parametrize the flat connections on $\CM_3$.
Moreover there are supersymmetric configurations with fluxes across two-cycles in $\CM_3$, which correspond to complex saddle points of the localizing action, because \eqref{eq:Dflux} is not satisfied along the initial contour for the auxiliary field $D$.\footnote{The inclusion of such complex saddles in the partition function has been discussed for instance in \cite{Benini:2016hjo,Closset:2016arn,Benini:2015noa,Honda:2015yha,Closset:2015rna,Gadde:2015wta}, in the context of localization on $T^n \times \Sigma_g$ ($n=0,1,2$) type manifold.} Below we will also write down the effective potential evaluated on the configurations with fluxes, and we will assume that we have to include the sum over the fluxes in the localization formula. Recall that a supersymmetric manifold $\CM_3$ is a circle fibration over a Riemann surface \cite{Closset:2012ru}. When the Riemann surface has genus $g > 1$, the flat connections on $\CM_3$ come in supersymmetric multiplets together with zero modes of the gauginos. Taking properly into account these zero modes, and the singularities that arise when matter multiplets become massless, leads to a particular choice of contour for the integral over the holonomies, that has been discussed in \cite{Benini:2013nda, Benini:2013xpa}. Here we will only be concerned with the form of the integrand, that arises from integrating out massive modes, and study it as a function of $a^i$ in the limit $\beta \to 0$. For this particular question, the dependence on the choice of $\CM_3$ is simple, and enters only through the densities \eqref{eq:AM3}, \eqref{eq:LM3} and \eqref{eq:lM3}. In fact, this limit is just obtained by adding to $V^{\rm eff} _{\CM_3}(a)$ in \eqref{eq:result} the contribution of the KK modes of the vector multiplet. This amounts to extending the sum in \eqref{eq:result} to include off-diagonal gauginos, i.e. $\rho_f$ is a root vector of $G$ and $R_f = 1$. Two observations are in order:
\begin{itemize}
\item{Even though $\kappa(x)$ and $\vartheta(x)$ are piecewise cubic and quadratic functions of $x$, the anomaly cancellation conditions \eqref{eq:anoodd} and \eqref{eq:anoeven} are equivalent to a cancellation of the terms of highest degree in $V^{\rm eff}_{\CM_3}(a)$. As a result, the $\kappa$ term is piece-wise quadratic in $a$ and the $\vartheta$ term is piece-wise linear;}
\item{The limit of the integrand is a function of $a$, and it can also depend on the fluxes and the additional holonomies of the gauge field on $\CM_3$ through the quantity $l_{\CM_3}^i$.}
\end{itemize}
We will now illustrate the formula by evaluating it in some examples, which we will then compare to localization.

%%%%%%%%%%%%%%%%%%%%%%%%%%%%%
\subsubsection{Example: $S^1 \times S^3_b$ and $S^1 \times L(n,1)$}
%%%%%%%%%%%%%%%%%%%%%%%%%%%%%%

Consider $S^1 \times S^3_b$ with the following squashed metric on $S^3$
\be
h_{\mu\nu}dx^\mu dx^\nu = r_3^2\left[b^{-2}\cos^2\psi d\phi^2 + b^2 \sin^2 \psi d\chi^2 + f(\psi)^2 d \psi^2\right]~.
\ee
In order to preserve supersymmetry we also need the following background fields
\be
H = - \frac{i}{r_3 f(\psi)}~,~~v_\mu = 0~.
\ee
The function $f(\psi)$ that appears in the metric and in $H$ is not constrained by supersymmetry, it is only required to satisfy $f(\tfrac\pi 2) = f(0)^{-1} = b$ to avoid singularities in the metric.

Since there are no nontrivial two-cycles in this space, the gauge field has $F^{i}_{\mu\nu} = 0$. The BPS condition implies also $D^{i} = 0$, and therefore $l^i_{S^3_b} = 0$. We have
\be
A_{S^3_b} = \frac{i}{\pi^2}\int_{S^3_b} d^3 x \sqrt{h} 2 H = 4 r_3^2 ~.
\ee
$L_{S^3_b}$ was already evaluated in \cite{DiPietro:2014bca}, the result is $L_{S^3_b} = 4 r_3 \frac{b + b^{-1}}{2}$. As a result we find the potential for the holonomies
\be
V^{\text{eff}}_{S^3_b}(a) =  -\sum_f \sum_{\rho_f \in \mathfrak{R}_f} \left(\frac{2 \pi^3 i r_3^2}{3 \beta^2}\kappa(\rho_f \cdot a) + \frac{2 \pi^2 r_3}{ \beta} \frac{b + b^{-1}}{2} R_f\vartheta(\rho_f \cdot a)   \right)~. \label{eq:squspheff}
\ee
This matches the result of \cite{Ardehali:2015bla}, that we will also reproduce in the next section. We note that both in $A_{\CM_3}$ and $L_{\CM_3}$ the dependence on the arbitrary function $f(\psi)$ cancels.

We can also consider the space $S^1\times L(n,1)$, where $L(n,1)$ is the Lens space. This space can be seen as $S^1 \times S^3/\mathbb{Z}_n$, where $\mathbb{Z}_n$ is a subgroup of the $U(1)$ isometry that rotates the Hopf fiber. The supergravity background fields have the same value as in $S^1\times S^3$, therefore the densities that we need to integrate to obtain $L_{L(n,1)}$ and $A_{L(n,1)}$ are the same as the ones in $L_{S^3}$ and $A_{S^3}$. The difference is that the integration now produces an additional factor of $1/n$, due to the reduced size of the Hopf fiber. Therefore we get
\be
V^{\text{eff}}_{L(n,1)}(a) = \frac 1 n V^{\text{eff}}_{S^3_b}(a)~.\label{eq:Lenseff}
\ee

%%%%%%%%%%%%%%%%%%%%%%%%%%%%%%
\subsubsection{Example: $T^2 \times \Sigma_g$}
%%%%%%%%%%%%%%%%%%%%%%%%%%%%%
Consider the product $T^2\times \Sigma_g$ of a torus and a Riemann surface of genus $g$. We can use coordinates $X$ and $Y$ on the torus, with identifications $X\sim X + \beta$ and $Y\sim Y + L$, and complex coordinates $(z,\bar{z})$ on $\Sigma_g$.

The background fields turned on in this background are the metric -- which we can take to be a hermitean metric on $\Sigma_g$ times a flat metric on the torus -- and the $R$-symmetry gauge field $A^{(R)}_M$, $M=1,\dots,4$. The latter is chosen in such a way to cancel the spin connection in the covariant derivative of the spinor on $\Sigma_g$
\be
A^{(R)}_M = - \frac 12 \omega_M^{23}~,
\ee 
where $2,3$ denote the flat indices on the Riemann surface. This implies that the $23$ component of the curvature two-form is proportional to the field strength of the $R$-symmetry $R^{23} = - 2 F^{(R)}$. Since
\be
\int_{\Sigma_g} R^{23} = \int_{\Sigma_{g}} {\rm Pf}_\Sigma(R)  = 4 \pi (1-g)~,
\ee
where $g$ is the genus, for $g\neq 1$ the Riemann surface supports also a flux of the $R$-symmetry gauge-field
\be
\int_{\Sigma_g} F^{(R)} = - 2 \pi (1-g)~.
\ee
i.e. a monopole background. Let us fix some notation about the complex coordinates on $\Sigma$. We define local holomorphic coordinates $(z, \bar{z})$, so that the non-zero components of the complex structure are $J^z_{~z} = i$ and $J^{\bar{z}}_{~\bar{z}} = -i$. We write the hermitean metric as $ds^2 = 2 g_{z\bar{z}}dz d\bar{z}$ and therefore $\sqrt{g} = 2 g_{z\bar{z}}$. The fields $v_\mu$ and $H$ vanish in this background. 

We can also consider configurations with a monopole background on $\Sigma_g$ for the background vector multiplet
\begin{align}
A & = A_X dX + A_Y dY + (A_z dz + c.c.)~,~\\
D & = -\frac12 J^{M N} F_{M N}  = - i g^{z\bar{z}}  F_{z\bar{z}}~~,
\end{align}
where $F^{i}_{z\bar{z}}$ has GNO-quantized flux on $\Sigma_g$
\be
\int_{\Sigma_g} F^{i} = 2\pi \mathfrak{m}^i~,~~\rho_f\cdot \mathfrak{m} \in \mathbb{Z}~.
\ee
The evaluation of $L_{S^1 \times \Sigma_g}$ and $l^i_{S^1 \times \Sigma_g}$ gives
\begin{align}
L_{S^1 \times \Sigma_g} & = \frac{ 1}{ \pi^2}\int_0^L dY \int_{\Sigma_g} dz d\bar{z}\sqrt{g} ~\frac 14 R = \frac 2 \pi L (1 -g)~,\\
l^i_{S^1 \times \Sigma_g} & = \frac{1}{\pi^2} \int_0^L d Y \int_{\Sigma_g} dz d \bar{z} (-i g^{z\bar{z}} F^{i}_{z\bar{z}}) = \frac{1}{\pi^2} L \int_{\Sigma_g} F^{i} = \frac{2}{\pi} L \mathfrak{m}^i~.
\end{align}
Plugging in the formula for the potential, we obtain
\be
V^{\text{eff}}_{S^1 \times \Sigma_g}(a, \mathfrak{m})= - \sum_f \sum_{\rho_f \in \mathfrak{R}_f} \frac{\pi L }{\beta}(R_f(1-g) - \rho_f \cdot \mathfrak{m}) \vartheta(\rho_f \cdot a) ~. \label{eq:Sigmaeff}
\ee
Note that in this case the localization formula will contain both an integral over the holonomies and a discrete sum over $\mathfrak{m}$. The result \eqref{eq:Sigmaeff} shows that each term in the sum over $\mathfrak{m}$ will have a different $\mathfrak{m}$-dependent leading contribution in the limit $\beta \to 0$.

%%%%%%%%%%%%%%%%%%%%%%%%%%%%%
\section{Comparison with Localization} \label{sec:loc}
%%%%%%%%%%%%%%%%%%%%%%%%%%%%
In this section,
we verify in various examples that
the refined Cardy formula \eqref{eq:result}
obtained in the previous section
agrees with the $\beta\rightarrow 0$ limit of the integrand of the localization matrix model. The explicit form of the integrand has been written down in closed form for $\CM_3 = S^3$, $L(n,1)$ and $S^1 \times \Sigma_g$. Special functions of different type enter in the one-loop determinant on $\CM_3 = S^3$ and $L(n,1)$ compared to the one on $S^1 \times \Sigma_g$. Nonetheless we will see that in both cases the limit is captured by the effective potential described in the previous section. We collected the definition of the various special functions involved and the formulas for their asymptotic behavior in the limit $\beta \to 0$ in the appendix \ref{app:SpF}. The case in which the torus fibration is nontrivial, and the Riemann surface has genus $g > 0$, has been discussed in \cite{Nishioka:2014zpa}, but to our knowledge the determinant over the massive modes has not been written down in a regularized closed form. In this case our result gives a prediction about the limit of the integrand.
%%%%%%%%%%%%%%%%%%%%%%%%%%%%%%%%%%%%%%%%%%%%%%%%%%
%%%%%%%%%%%%%%%%%%%%%%%%%%%%%%%%%%%%%%%%%%%%%%%%%%
\subsection{Primary Hopf Surface}
%%%%%%%%%%%%%%%%%%%%%%%%%%%%%%%%%%%%%%%%%%%%%%%%%%
%%%%%%%%%%%%%%%%%%%%%%%%%%%%%%%%%%%%%%%%%%%%%%%%%%
Let us consider 
the partition function on a primary Hopf surface.
A primary Hopf surface is defined as the following identification\footnote{
Though there is another type of primary Hopf surface
given by the identification $(z_1 ,z_2 )\sim (q^n z_1 +\lambda z_2^n ,qz)$
with $\lambda\in\mathbb{C}$ and $n\in\mathbb{N}$,
we do not consider this type.
} 
of $\mathbb{C}^2-(0,0)$ with coordinates $(z_1 ,z_2 )$:
\begin{\eq}
(z_1 ,z_2 ) \sim (p z_1 , q z_2 )~. 
\label{eq:defHopf}
\end{\eq}
The choice of parameters $(p,q)$ determines the complex structure. 
Here we will restrict to
\begin{\eq}
p=e^{2\pi i\sigma} =e^{-\frac{\beta}{r_3} b_1} ,\quad
q=e^{2\pi i\tau} =e^{-\frac{\beta}{r_3} b_2} \quad 
(b_1 ,b_2 \in\mathbb{R}_+ )~.
\end{\eq}
The topology of this space is $S^1 \times S^3$ \cite{Kodaira}.
When $b_1 =b_2^{-1}=b$
this space becomes $S^1 \times S^3_b$ and
the effective potential was obtained in \cite{Ardehali:2015bla}. The derivation presented there straightforwardly carries over to generic $(b_1,b_2)$ and we will now review it.

We can exactly compute
the partition function of a 4d $\mathcal{N}=1$ theory on this space
by using localization.
The saddle point configuration of localization is given by
flat connections of the gauge field and zero for all the other fields.
The flat connection is parametrized 
by holonomies along $S^1$.
The partition function is given by \cite{Assel:2014paa}
\be
Z_{\rm Hopf}
=  e^{-\beta E_{\rm susy}}\frac{(p;p)^r (q;q)^r}{|W|} \underset{-\half \leq a^i \leq \half}{\int} d^r a \
 Z_{\rm vec}^{\rm Hopf} Z_{\rm chi}^{\rm Hopf}~,
\ee
where $r$ is the rank of the gauge group $G$, $|W|$ is the order of the Weyl group of $G$, $E_{\rm susy}$ is the supersymmetric Casimir energy \cite{Assel:2015nca}, and $(x;x) = \displaystyle{\prod_{k= 0}^{\infty}}(1-x^{k+1})$. 
$Z^{\rm Hopf}_{\rm vec}$ 
is the contribution from the one-loop determinant of the vector multiplet, given by
\begin{\eq}
Z_{\rm vec}^{\rm Hopf}
= \frac{1}{\displaystyle{\prod_{\alpha\in \Delta} } \Gamma_e ( e^{2\pi i\alpha\cdot a} ; p ,q ) }~,
\end{\eq}
where $\Delta$ is the set of roots, and $\Gamma_e (e^{2\pi ia};e^{2\pi i\sigma} ,e^{2\pi i\tau} ) \equiv \Gamma (a;\sigma ,\tau )$ 
is the elliptic gamma function \eqref{eq:EllG}. The one-loop determinant $Z^{\rm Hopf}_{\rm chi}$ of the chiral multiplet is
\begin{\eq}
Z_{\rm chi}^{\rm Hopf}
=\prod_I \prod_{\rho_I \in \mathfrak{R}_I} 
 \Gamma_e ((pq)^{R_I /2} e^{2\pi i \rho_I \cdot a} ; p ,q )~, 
\end{\eq}
where $I$ runs over all the matter multiplets, $\mathfrak{R}_I$ denotes their representation under $G$, with associated weight $\rho_I$, and $R_I$ is the $R$-charge of the scalar in the chiral multiplet.

Now let us see the $\beta\rightarrow 0$ behavior of the one-loop determinant. Using \eqref{eq:LimEllG}, we find
\begin{align}
& \ln Z_{\rm vec}^{\rm Hopf}
\underset{\beta \to 0}{\longrightarrow} 2\pi i \frac{\tau +\sigma}{4\tau\sigma}
\sum_{\alpha \in \Delta}  \left( \vartheta (\alpha\cdot a) -\frac{1}{6}\right) +\mathcal{O}(\beta^0 )~, \\
& \ln Z_{\rm chi}^{\rm Hopf}
\underset{\beta \to 0}{\longrightarrow} 2\pi i \sum_I \sum_{\rho_I \in \mathfrak{R}_I}
 \Biggl[ -\frac{\kappa (\rho_I \cdot a)}{12\tau\sigma}
+(R_I -1) \frac{\tau +\sigma}{4\tau\sigma} \left( \vartheta (\rho_I \cdot a) -\frac{1}{6}\right) \Biggr] +\mathcal{O}(\beta^0 )~.
\end{align}
Thus
the total one-loop determinant in the $\beta\rightarrow 0$ limit is
\begin{align}
& Z_{\rm vec}^{\rm Hopf} Z_{\rm chi}^{\rm Hopf}
\underset{\beta \to 0}{\longrightarrow} 
\displaystyle{e^{-\frac{\pi^2}{3\beta} \frac{b_1 +b_2}{2b_1 b_2} {\rm Tr}(R)}
e^{-V^{\rm eff}_{\rm Hopf}(a) + \CO(\beta^0)} }~,\nonumber \\
& V^{\rm eff}_{\rm Hopf}(a)  = -\frac{2\pi^3 i r_3^2 }{3 b_1 b_2 \beta^2}\sum_I \sum_{\rho_I} \kappa (\rho_I \cdot a ) \nonumber\\& -\frac{2\pi^2 r_3}{\beta} \frac{b_1 +b_2}{2b_1 b_2} \left(\sum_I (R_I -1 )\sum_{\rho_I\in \mathfrak{R}_I}\vartheta (\rho_I \cdot a)
+ \sum_{\alpha \in \Delta}\vartheta (\alpha \cdot a)\right)~.\label{eq:Hopflocres}
\end{align}
Recall that $R_I -1$ is the $R$-charge of the fermion in the chiral multiplet, and the $R$-charge of gauginos is 1. Since the roots can be split into positive and negative, and $\kappa(-x) = -\kappa(x)$, we have
\be
\sum_{\alpha \in \Delta} \kappa(\alpha\cdot a) = 0~,
\ee
so that we can freely extend the sum in the first line to include also the (vanishing) contribution of gauginos. Therefore matter fermions and gauginos contribute in the same way to the final result. For $b_1 = b_2 = b^{-1}$ this is the result of \cite{Ardehali:2015bla}, and as anticipated it matches with the effective theory calculation \eqref{eq:squspheff}.

%%%%%%%%%%%%%%%%%%%%%%%%%%%%%%%%%%%%%%%%%%%%%%%%%%
%%%%%%%%%%%%%%%%%%%%%%%%%%%%%%%%%%%%%%%%%%%%%%%%%%
\subsection{$S^1 \times L(n,1)$}
%%%%%%%%%%%%%%%%%%%%%%%%%%%%%%%%%%%%%%%%%%%%%%%%%%
%%%%%%%%%%%%%%%%%%%%%%%%%%%%%%%%%%%%%%%%%%%%%%%%%%
Next we
consider 
the partition function on $S^1_\beta \times L(n,1) \simeq S^1_\beta \times S^3_b /\mathbb{Z}_n$, which is given by the following further identification of \eqref{eq:defHopf} 
with $b_1 =b=b_2^{-1}$ 
\begin{\eq}
(z_1 ,z_2 ) \sim (w_n z_1 , w_n^{-1}z_2 ) ,
\end{\eq}
where $w_n =e^{\frac{2\pi i}{n}}$. The Cardy behavior of the integrand for this partition function was also considered in \cite{Nieri:2015yia}.

The saddle point configuration of localization is given by
flat connections of the gauge field and zero for all the other fields.
The flat connection is parametrized 
by the holonomies along $S^1$
and along the non-contractible cycle of $L(n,1)$. 
Since $\pi_1 (L(n,1)) =\mathbb{Z}_n$, the latter are
specified by a set of integers $(m_1 ,\cdots ,m_r)$
with $0\leq m_i \leq n-1$.
The partition function is given by \cite{Benini:2011nc,Razamat:2013opa}
(see also \cite{Kinney:2005ej,Romelsberger:2005eg,Dolan:2008qi,Nawata:2011un,Imamura:2011uw,Assel:2014paa,Nishioka:2014zpa})\footnote{
There is an additional factor independent of $a$
but this is $\exp{(\mathcal{O}(\beta ))}$ and irrelevant for our purpose.
}
\begin{\eq}
Z_{S^1 \times L(n,1)}
= \sum_{\{m_i \}=0}^{n-1} \frac{1}{n^r|W_{\{m_i\}}|} \underset{-\half \leq a_i \leq \half}{\int} d^r a \
 Z_{{\rm vec}\,\{m_i\}}^{S^1 \times L(n,1)} 
Z_{{\rm chi}\,\{m_i\} }^{S^1 \times L(n,1)},
\end{\eq}
where $|W_{\{m_i\}}|$ is the rank of Weyl group of residual gauge group,
and
\begin{align}
& Z_{{\rm vec}\,\{m_i\}}^{S^1 \times L(n,1)}
= \frac{1}{\displaystyle{\prod_{\alpha\in \Delta}} 
\Gamma_e ( e^{2\pi i \alpha\cdot a} p^{\alpha\cdot m}; pq ,p^n ) 
\Gamma_e (e^{2\pi i \alpha\cdot a} q^{n-\alpha\cdot m}; pq ,q^n )}~, \\
& Z_{{\rm chi}\,\{m_i\}}^{S^1 \times L(n,1)}
=\prod_I \prod_{\rho_I \in {\mathfrak{R}_I}} 
\Gamma_e ((pq)^{\frac{R_I}{2}} e^{2\pi i \rho_I \cdot a} p^{\rho_I \cdot m}; pq ,p^n ) 
\Gamma_e ((pq)^{\frac{R_I}{2}} e^{2\pi i \rho_I \cdot a}  q^{n-\rho_I \cdot m}; pq ,q^n )~.\nonumber
\end{align}

Now let us see the $\beta\rightarrow 0$ behavior of the one-loop determinant.
Using \eqref{eq:LimEllGn1}-\eqref{eq:LimEllGn2} we find
\begin{align}
& \ln Z_{{\rm vec}\,\{m_i\}}^{S^1 \times L(n,1)}
\underset{\beta \to 0}{\longrightarrow}  \frac{2\pi i}{n} 
\frac{\tau +\sigma}{4\tau\sigma}
\sum_{\alpha \in \Delta}  \left( \vartheta (\alpha\cdot a) -\frac{1}{6}\right) +\mathcal{O}(\beta^0 )~, \nonumber\\ 
& \ln Z_{{\rm chi}\,\{m_i\}}^{S^1 \times L(n,1)}
\underset{\beta \to 0}{\longrightarrow}   \frac{2\pi i}{n} \sum_I \sum_{\rho_I \in {\mathfrak{R}_I}}
 \Biggl[ -\frac{\kappa (\rho_I \cdot a)}{12\tau\sigma}
+(R_I -1) \frac{\tau +\sigma}{4\tau\sigma} \left( \vartheta (\rho_I \cdot a) -\frac{1}{6}\right) \Biggr] +\mathcal{O}(\beta^0 )~. \nonumber\\
\end{align}
Note that remarkably the dependence on $m_i$ cancels in the leading terms for $\beta\to 0$. Thus
the total one-loop determinant in the $\beta\rightarrow 0$ limit is
\begin{align}
 \sum_{\{m_i \}=0}^{n-1} Z_{{\rm vec}\,\{m_i\}}^{S^1 \times L(n,1)}
  Z_{{\rm chi}\,\{m_i\}}^{S^1 \times L(n,1)} 
&  \underset{\beta \to 0}{\longrightarrow} 
\displaystyle{e^{-\frac{1}{n}
\frac{\pi^2 r_3}{3\beta} \frac{b +b^{-1}}{2} {\rm Tr}(R)}e^{- V^{\rm eff}_{L(n,1)}(a) + \mathcal{O}(\beta)} }~, \nonumber\\
V^{\rm eff}_{L(n,1)}(a) & =\frac{1}{n}V^{\rm eff}_{S^3_b}(a)~.
\end{align}
Every term in the sum over $m_i$ contributes the same, and the 
$n$-dependence appears only as an overall factor. The result agrees with the one derived via the effective theory approach \eqref{eq:Lenseff}.
%%%%%%%%%%%%%%%%%%%%%%%%%%%%%%%%%%%%%%%%%%%%%%%%%%
%%%%%%%%%%%%%%%%%%%%%%%%%%%%%%%%%%%%%%%%%%%%%%%%%%
\subsection{$T^2 \times \Sigma_g$}
%%%%%%%%%%%%%%%%%%%%%%%%%%%%%%%%%%%%%%%%%%%%%%%%%%
%%%%%%%%%%%%%%%%%%%%%%%%%%%%%%%%%%%%%%%%%%%%%%%%%%
Next we consider partition function on $T^2 \times \Sigma_g$,
where $\Sigma_g$ is Riemann surface with genus $g$.
The localization computation was presented in \cite{Benini:2016hjo}
(see also \cite{Closset:2016arn,Benini:2015noa,Honda:2015yha,Closset:2015rna}).
The saddle point configuration of  localization is
somewhat involved on $T^2 \times \Sigma_g$.
Along the original integral contour of the path integral
this is simply given by flat connections on $T^2 \times \Sigma_g$ 
and zero modes.
In addition, there are complex saddle points with flux along $\Sigma_g$,
which may contribute to the path integral.
Below we perform the $\beta \to 0$ limit of the localization formula
assuming that the complex saddle points contribute to the path integral.

We take coordinate $X$ and $Y$ on the torus, with $X \sim X + \beta$, $Y \sim Y + L$. The flat connection on $T^2$ is parametrized as
\be
u 
= \frac {1}{2\pi}\left(\int_0^{\beta}d X A_X - \tau\int_0^L dY A_Y\right)~,~~
\bar{u} 
=  \frac {1}{2\pi}\left(\int_0^{\beta} dX A_X -\bar{\tau}\int_0^LdY A_Y\right)~,
\ee
where $\tau$ is the complex structure of $T^2$. After integration over $\bar{u}$, taking into account the gaugino zero-modes,
the localization formula of the partition function is given by
\begin{\eq}
Z_{T^2 \times \Sigma_g}
=\sum_{\{\mathfrak{m}^i \}} \frac{1}{|W|} \oint_C d^ru\ 
Z_{{\rm 1loop}\,\{ \mathfrak{m}^i \}}^{T^2 \times \Sigma_g}
\Biggl[ {\rm det}_{ij}\left( 
\frac{\del^2}{\del (iu^i)\del \mathfrak{m}^j} \log{Z_{{\rm 1loop}\,\{ \mathfrak{m}^i \}}^{T^2 \times \Sigma_g}}
 \right) \Biggr]^g ,
\end{\eq}
where $\{ \mathfrak{m}^i \}$ is the set of GNO quantized fluxes of $U(1)^r$ across $\Sigma_g$.
The choice of contour $C$ is complicated\footnote{$C$ is chosen 
to give the Jeffrey-Kirwan residue \cite{JK,Benini:2013xpa}
at some singularities of the integrand.
} 
but it is irrelevant for our purpose.
The one-loop determinant $Z_{{\rm 1loop}\,\{ \mathfrak{m}^i \}}^{T^2 \times \Sigma_g}$
is the product of the contributions from vector and chiral multiplets
\begin{align}
Z_{{\rm 1loop}\,\{ \mathfrak{m}^i \}}^{T^2 \times \Sigma_g} (\tau ,u)
&= Z_{{\rm vec}\,\{ \mathfrak{m}^i \}}^{T^2 \times \Sigma_g} (\tau ,u ) 
Z_{{\rm chi}\,\{ \mathfrak{m}^i \}}^{T^2 \times \Sigma_g} (\tau ,u )~,\NN\\
Z_{{\rm vec}\,\{ \mathfrak{m}^i \}}^{T^2 \times \Sigma_g} (\tau ,u )
&= \left( \frac{2\pi\eta^2 (\tau )}{i}\right)^r
  \prod_{\alpha\in \Delta} 
 \left( \frac{i\theta_1 (\tau |\alpha\cdot u )}{\eta (\tau )} \right)^{1-g}~,\\
Z_{{\rm chi}\,\{ \mathfrak{m}^i \}}^{T^2 \times \Sigma_g} (\tau ,u )
&= \prod_I \prod_{\rho_I \in \mathfrak{R}_I} 
 \left( \frac{i\eta (\tau )}{\theta_1 (\tau |\rho_I \cdot u )} 
\right)^{\rho_I \cdot \mathfrak{m} +(g-1)(R_I -1)}~,
\end{align}
where $\eta (\tau )$ is Dedekind eta function and
$\theta_1 (\tau |z)$ is Jacobi theta function, see \eqref{eq:EtaTheta}.

Now we are interested in the limit $\beta\to 0$ with $L$ fixed.
Taking 
$\tau =i{\rm Im}\tau =i \beta /L$ we have $u = \beta ( A_X -i A_Y )$
with $A_X \in [-\pi / \beta,\pi /\beta ] ,A_Y \in [-\pi / L,\pi /L ]$.
Hence,
keeping $a =\beta A_X /2\pi$ fixed in the limit,
the variable $u$ projects to
\begin{\eq}
u = a -i\frac{\beta}{2\pi} A_Y \underset{\beta \to 0}{\longrightarrow}  a~.
\end{\eq}

Now let us write the $\beta\rightarrow 0$ behavior of the one-loop determinant. Using \eqref{eq:EtaLim}-\eqref{eq:ThetaLim}, we obtain
\begin{align}
\log Z_{{\rm vec}\,\{\mathfrak{m}^i\}}^{T^2 \times \Sigma_g} (\tau ,u )
&\underset{\beta \to 0}{\longrightarrow} 
\frac{\pi L }{\beta}\sum_{\alpha\in \Delta} 
(1-g)  \left(\vartheta(\alpha\cdot a) - \frac 16 \right) + \mathcal{O}(\log \beta)~, \label{eq:veclimT2Sigma} \\
\log Z_{{\rm chi}\,\{\mathfrak{m}^i\}}^{T^2 \times \Sigma_g} (\tau ,u)
&\underset{\beta \to 0}{\longrightarrow} \frac{\pi L }{\beta} \sum_I \sum_{\rho_I \in \mathfrak{R}_I} 
(-\rho_I \cdot \mathfrak{m} +(1-g)(R_I -1))\left(\vartheta(\rho_I \cdot a) - \frac 16\right)\nonumber \\ 
&~~~~~~~+ \,\mathcal{O}(\log \beta)~. 
\end{align}

Since the ${\rm det}_{ij}$ factor contributes by $\mathcal{O}(\log \beta )$,
the limit of the integrand in the $T^2 \times \Sigma_g$ partition function 
is dominated by the limit of the 1-loop determinants above, namely
\begin{align}
 & Z_{{\rm vec}\,\{\mathfrak{m}^i\}}^{T^2 \times \Sigma_g}  Z_{{\rm chi}\,\{\mathfrak{m}^i\}}^{T^2 \times \Sigma_g} \underset{\beta \to 0}{\longrightarrow} \displaystyle{e^{-(1-g)\frac{\pi L}{6 \beta} \Tr(R)}  e^{- V^{\rm eff}_{S^1 \times \Sigma_g}(a, \mathfrak{m}) + \mathcal{O}(\log \beta)}}~, \nonumber\\
& V^{\rm eff}_{S^1 \times \Sigma_g}(a, \mathfrak{m}) 
= - \frac{\pi L}{\beta} \left(\sum_{\alpha \in \Delta} (1-g)\vartheta(\alpha\cdot a) \right. \nonumber \\
&~~~~~~~~~~~~~~~~~~~~~~~~~~ 
+ \left. \sum_I \sum_{\rho_I \in \mathfrak{R}_I} (-\rho_I \cdot \mathfrak{m} + (1-g)(R_I -1))\vartheta(\rho_I \cdot a)\right)~.
\label{eq:locresT2Sigma}
\end{align}
Note that since $\vartheta(x) = \vartheta(-x)$, and the roots can be split in positive and negative, for any given $\mathfrak{m}$ we have 
\be\displaystyle{\sum_{\alpha\in\Delta} }\alpha\cdot \mathfrak{m}\,( \vartheta(\alpha\cdot a) - \tfrac 1 6 ) = 0~. 
\ee
Therefore we can freely add a $\alpha\cdot \mathfrak{m}$ term in the sum over the roots in \eqref{eq:locresT2Sigma}, so that the contributions of vector and chiral multiplets take the same form. In the $a$-independent prefactor we have used the gauge-gravitational-gravitational anomaly cancellation
condition
\be
\sum_I \sum_{\rho_I\in \mathfrak{R}_I} \rho_I^i =0~.
\ee
The result \eqref{eq:locresT2Sigma} agrees with the effective theory calculation \eqref{eq:Sigmaeff}. 

%%%%%%%%%%%%%%%%%%%%%%%%%%%%%%%%%%%%%%%%%%%%%%%%%%
%%%%%%%%%%%%%%%%%%%%%%%%%%%%%%%%%%%%%%%%%%%%%%%%%%
%%%%%%%%%%%%%%%%%%%%%%%%%%%%%%%%%%%%%%%%%%%%%%%%%%
\section{Minima of $V^{\rm eff}_{\CM_3}$ and $Z_{\CM_3}$}\label{sec:min}
%%%%%%%%%%%%%%%%%%%%%%%%%%%%%%%%%%%%%%%%%%%%%%%%%%
%%%%%%%%%%%%%%%%%%%%%%%%%%%%%%%%%%%%%%%%%%%%%%%%%%
%%%%%%%%%%%%%%%%%%%%%%%%%%%%%%%%%%%%%%%%%%%%%%%%%%
In \cite{Ardehali:2015bla}
Arabi Ardehali 
found that
``high temperature" limit of the superconformal index behaves as
\begin{\eq}
Z_{S^1 \times S^3_b}
\underset{\beta \to 0}{\longrightarrow} e^{-\frac{\pi^2 r_3}{3\beta}\frac{b+b^{-1}}{2}{\rm Tr}(R)}
\int d^r a\ 
e^{-V^{\rm eff}_{S^3_b}(a)}~,
\end{\eq}
with $V^{\rm eff}_{S^3_b}(a)$ as in \eqref{eq:squspheff}.
Naively, since the exponent becomes very large for $\beta\rightarrow 0$,
the limit of the integral should be controlled by the minimum of the effective potential. This was shown rigorously to be true in \cite{Ardehali:2015bla}, at least for theories with charge conjugation invariance. Thus, if the minimum of the effective potential is not zero, the supersymmetric Cardy formula \eqref{eq:Cardy} receives corrections.

Arabi Ardehali showed that when the potential has a minimum in the origin the localization integrand for $Z_{S^3_b}$ decreases exponentially at infinity. Vice versa, if the origin is a local maximum, and there is a nontrivial minimum $V^{\rm eff}_{S^3_b}(a_{\rm min}) < 0$, then the localization integrand grows exponentially at infinity. In this section we will extend this result and show that
the same relation holds between the minimum of $V^{\rm eff}_{\CM_3}(a)$,
and the integrand of the localization formula for the 3d partition function, 
for the following geometries: Hopf surfaces,
$S^1 \times L(n,1)$ and $T^2 \times \Sigma_g$.

%%%%%%%%%%%%%%%%%%%%%%%%%%%%%%%%%%%%%%%%%%%%%%%%%%
%%%%%%%%%%%%%%%%%%%%%%%%%%%%%%%%%%%%%%%%%%%%%%%%%%
\subsection{Primary Hopf Surface and $S^1 \times L(n,1)$}
%%%%%%%%%%%%%%%%%%%%%%%%%%%%%%%%%%%%%%%%%%%%%%%%%%
%%%%%%%%%%%%%%%%%%%%%%%%%%%%%%%%%%%%%%%%%%%%%%%%%%
%%%%%%%%%%%%%%%%%%%%%%%%%%%%%%%%%%%%%%%%%%%%%%%%%%
\subsubsection{Condition for minimality of $V^{\rm eff}(a=0)=0$}
%%%%%%%%%%%%%%%%%%%%%%%%%%%%%%%%%%%%%%%%%%%%%%%%%%
The real part of the effective potential 
\eqref{eq:Hopflocres} for Hopf surface is 
\be
\Re(V^{\rm eff}_{\rm Hopf}) (a)
\equiv- \frac{2\pi^2 r_3}{\beta} \frac{b_1 +b_2}{2b_1 b_2}\left(\sum_I (R_I-1 ) \sum_{\rho_I \in \mathfrak{R}_I} \vartheta ( \rho_I \cdot a ) 
 +\sum_{\alpha \in \Delta } \vartheta ( \alpha \cdot a ) \right)~.
\ee
Note that $\Re(V^{\rm eff}_{\rm Hopf})(a=0) =0$. Now let us ask
when $\Re(V^{\rm eff}_{\rm Hopf})(a=0) =0$ is a local minimum or maximum. We can study this by looking at the following quantity
\begin{\eq}
\lim_{|a|\rightarrow +0} \frac{\del\Re(V^{\rm eff}_{\rm Hopf})(a)}{\del |a|} .
\end{\eq}
If this quantity is positive (negative) for all the directions $a^i$, $i = 1, \dots, r$,
then $\Re(V^{\rm eff}_{\rm Hopf})(a=0)$ $=0$ is local minimum (maximum).

For this purpose,
it is sufficient to take a region of $a$ such that $|\alpha\cdot a|<1$ and $|\rho_I \cdot a|<1$ for every $\rho_I$. 
Since $\vartheta (x)=|x|-x^2$ for $|x|<1$,
we obtain
\begin{\eqa}
\frac{\del \Re(V^{\rm eff}_{\rm Hopf}) (a)}{\del |a|} 
\propto - \frac{1}{|a|}\Biggl[ \sum_I (R_I -1 )  \sum_{\rho_I \in \mathfrak{R}_I} |\rho_I \cdot a |  
 + \sum_{\alpha \in \Delta } |\alpha\cdot a  | \Biggr]~,
\label{eq:dLh}
\end{\eqa}
where we have used $U(1)_R$-gauge-gauge anomaly cancellation condition:
\begin{\eq}
\sum_I (R_I -1 )  \sum_{\rho_I \in \mathfrak{R}_I} \rho_I^i \rho_I^j
 +\sum_{\alpha \in \Delta } \alpha^i \alpha^j =0~. 
\end{\eq}
This means that 
local minimality of $V^{\rm eff}_{\rm Hopf}(a = 0)=0$ is determined by the sign of the following quantity
\be
\tilde{L}_{\rm Hopf} (a)
= - \frac{1}{2}\sum_I (R_I -1 )  \sum_{\rho_I \in \mathfrak{R}_I} |\rho_I \cdot a |  
 -\half \sum_{\alpha \in \Delta } |\alpha \cdot a  |~.\label{eq:LminHopf}
\ee
If $\tilde{L}_{\rm Hopf} (a)>0$ ($<0$), 
then $V^{\rm eff}_{\rm Hopf}(a = 0)=0$ is a local minimum (maximum). 
Since \eqref{eq:dLh} is valid 
in the range of $|\alpha\cdot a|<1$ and $|\rho_I \cdot a|<1$,
$V^{\rm eff}_{\rm Hopf}(a = 0)=0$ is a minimum (maximum) in this region
when $\tilde{L}_{\rm Hopf} (a)>0$ ($<0$), 
and we cannot draw conclusions on the global minimum (maximum). 
On the other hand, $\tilde{L}_{\rm Hopf} > 0$ ($<0$) in the neighborhood of the origin implies $\tilde{L}_{\rm Hopf} > 0$ ($<0$) for any $a$.
The above results hold also for the Lens index
since $V^{\rm eff}_{L(n,1)}(a)=V^{\rm eff}_{\rm Hopf}(a) /n$.

%%%%%%%%%%%%%%%%%%%%%%%
\subsubsection{Behavior at infinity of the $Z_{S^3_b}$ integrand}
%%%%%%%%%%%%%%%%%%%%%%%
Let us consider partition function on the squashed sphere $S^3_b$.
The saddle point configuration of localization for this space is given by 
flat connections of the gauge fields and 
a constant configuration of the adjoint scalars in the vector multiplet.
The localization formula is given by\footnote{
This formula is for theories without Chern-Simons terms, FI terms and real masses.
} \cite{Hama:2011ea,Imamura:2011wg,Alday:2013lba,Kapustin:2009kz,Jafferis:2010un,Hama:2010av}
\be
Z_{S^3_b}
=\frac{1}{|W|} \int d^r \sigma \  Z^{S^3_b}_{\rm vec} Z^{S^3_b}_{\rm chi}~,
\ee
where
\begin{align}
& Z^{S^3_b}_{\rm vec}
=\prod_{\alpha\in {\rm root_+}} 4\sinh{(\pi b\alpha\cdot\sigma )}
\sinh{(\pi b^{-1}\alpha\cdot\sigma )}~,\\
& Z^{S^3_b}_{\rm chi}
=  \frac{1}{\displaystyle{\prod_I \prod_{\rho_I\in \mathfrak{R}_I} 
s_b\left( \rho_I \cdot \sigma  
    - i \frac{b + b^{-1}}{2}(1-R_I) \right) }}~. 
\end{align}
$s_b(z)$ is defined in \eqref{eq:DouSin}.
Now let us see the large-$\sigma$ behavior of the one-loop determinant.
By using \eqref{eq:DouSinLim} we find
\begin{align}
Z_{\rm vec}^{S^3_b}
&\underset{|\sigma| \to \infty}{\longrightarrow}\exp{\Biggl[ 
\frac{\pi (b + b^{-1})}{2}
\sum_{\alpha\in \Delta}|\alpha\cdot \sigma | \Biggr]}~, \\
Z_{\rm chi}^{S^3_b}
&\underset{|\sigma| \to \infty}{\longrightarrow}  \exp{\Biggl[  \sum_I \sum_{\rho_I\in \mathfrak{R}_I}
 \left( - \frac{i\pi {\rm sgn}(\rho_I \cdot\sigma )}{2}
   (\rho_I \cdot\sigma)^2 
-\frac{\pi (b+b^{-1})(1- R_I )}{2}|\rho_I \cdot\sigma |  
\right) \Biggr] }~. \nonumber
\end{align}
Therefore the whole integrand behaves as
\begin{\eq}
Z_{\rm vec}^{S^3_b} Z_{\rm chi}^{S^3_b}
\underset{|\sigma| \to \infty}{\longrightarrow}  \exp{\Biggl[  \left(
 -\sum_I \sum_{\rho_I} 
 \frac{i\pi {\rm sgn}(\rho_I \cdot\sigma )}{2} (\rho_I \cdot\sigma)^2 
-\pi (b + b^{-1}) \tilde{L}_{\rm Hopf} (\sigma ) 
\right) \Biggr] }~,
\label{eq:infinityS3}
\end{\eq}
where $\tilde{L}_{\rm Hopf}$ is the same as in \eqref{eq:LminHopf}. Clearly the integral is convergent (divergent) if $\tilde{L}_{\rm Hopf} > 0$ ($<0$).
We see that the existence of a local minimum in the origin 
for $V^{\rm eff}_{\rm Hopf}$ and 
the convergence of $Z_{S^3_b}$ are controlled by the sign of the same quantity, 
therefore they are equivalent conditions. 

Interestingly \eqref{eq:infinityS3} has a physical interpretation.
Since $\sigma$ is a dynamical version of a real mass parameter,
we can regard \eqref{eq:infinityS3}  as the effective theory after integrating out massive fields charged under $U(1)^r$.
From this point of view, the first term can be regarded as effective gauge-gauge Chern-Simons level, while the second term corresponds to effective FI parameter.
The effective FI parameter $\tilde{L}_{\rm Hopf}(\sigma)$ can be identified with
the $U(1)_R$ charge of the Coulomb branch operator associated to $\sigma$ \cite{Safdi:2012re,Lee:2016zud}.

%%%%%%%%%%%%%%%%%%%%%%%
\subsubsection{Behavior at infinity of the $Z_{L(n,1)}$ integrand}
%%%%%%%%%%%%%%%%%%%%%%%
One can exactly compute the partition function of
3d $\mathcal{N}=2$ SUSY theory on $L(n,1)=S^3_b /\mathbb{Z}_n$ using localization.
The saddle point configuration for this space is given by
flat connections of the gauge field and 
constant configurations of the adjoint scalar in the vector multiplet.
The possible values of flat connections are determined 
by $\pi_1 (L(n,1) ) =\mathbb{Z}_n$.
Therefore the localization formula is given 
by \cite{Imamura:2012rq,Alday:2012au,Benini:2011nc,Gang:2009wy}
\be
Z_{L(n,1)}
= \sum_{\{m_i \}=0}^{n-1} \frac{1}{n^r |W_{\{m^i\}}|} \int d^r \sigma 
 Z_{{\rm vec}\,\{m_i\}}^{L(n,1)} Z_{{\rm chi}\,\{m_i\}}^{L(n,1)}~.
\ee
The one-loop determinants are
\begin{align}
Z_{{\rm vec}\,\{m_i\}}^{L(n,1)} & =\prod_{\alpha\in \Delta} 
s_{b,\alpha\cdot m}\left( \alpha \cdot\sigma -i(b + b^{-1})/2 \right)~,\nonumber \\
Z_{{\rm chi}\,\{m_i\}}^{L(n,1)} & =  \frac{1}{\displaystyle{\prod_I \prod_{\rho_I\in \mathfrak{R}_I}} s_{b,\rho_I \cdot m}\left( \rho_I \cdot\sigma -i(b + b^{-1})(1-R_I )/2 \right)}~, 
\end{align}
where
\be
 s_{b,h}(z) 
= \prod_{k=0}^{n-1} s_b \left( \frac{z}{n} 
+ib\langle k\rangle_n  +ib^{-1}\langle k+h\rangle_n \right)~,~~
\langle m \rangle_n
\equiv \frac{1}{n}\left( [m]_n +\frac{1}{2} \right) -\frac{1}{2}~.
\ee
Now let us see large-$\sigma$ behavior of the one-loop determinant.
By using \eqref{eq:DouSinLim},
we find that the whole integrand behaves as
\begin{\eq}
\sum_{\{m_i \}=0}^{n-1} Z_{{\rm vec}\{m_i\}}^{L(n,1)} Z_{{\rm chi}\{m_i\}}^{L(n,1)}
\underset{|\sigma| \to \infty}{\longrightarrow}  \exp{\Biggl[ \frac{1}{n} \left(
 -\sum_I \sum_{\rho_I \in \mathfrak{R}_I} 
 \frac{i\pi {\rm sgn}(\rho_I \cdot\sigma )}{2} (\rho_I \cdot\sigma)^2 
-\pi (b + b^{-1}) \tilde{L}_{\rm Hopf} (\sigma ) 
\right) \Biggr] }~.
\label{eq:infinityLn}
\end{\eq}
The dependence on $\{m_i\}$ cancels in the limit, and every term in the sum contributes the same. $n$ appears only as an overall factor. We see that the convergence of the $Z_{L(n,1)}$ localization formula is
controlled by the sign of $\tilde{L}_{\rm Hopf} (\sigma )$. This again matches with the condition for $V^{\rm eff}_{L(n,1)}$ in \eqref{eq:Lenseff} to have a local minimum in the origin.

%%%%%%%%%%%%%%%%%%%%%%%%%%%%%%%%%%%%%%%%%%%%%%%%%%
%%%%%%%%%%%%%%%%%%%%%%%%%%%%%%%%%%%%%%%%%%%%%%%%%%
\subsection{$T^2 \times \Sigma_g$}
%%%%%%%%%%%%%%%%%%%%%%%%%%%%%%%%%%%%%%%%%%%%%%%%%%
%%%%%%%%%%%%%%%%%%%%%%%%%%%%%%%%%%%%%%%%%%%%%%%%%%
%%%%%%%%%%%%%%%%%%%%%%%%%%%%%%%%%%%%%%%%%%%%%%%%%%
\subsubsection{Condition for minimality of $V^{\rm eff}(a=0)= 0$}
%%%%%%%%%%%%%%%%%%%%%%%%%%%%%%%%%%%%%%%%%%%%%%%%%%
As for $S^1 \times S^3_b$, we impose the local minimum in $a = 0$ by taking
derivatives of the potential with respect to $|a|$ 
in the region of $|\rho_I \cdot a|<1$ and $|\alpha\cdot a|<1$. Using the formula \eqref{eq:Sigmaeff} for $V^{\rm eff}_{S^1 \times \Sigma_g}$ (in this case the potential is real), we obtain
\be
\frac{\del V^{\rm eff}_{S^1\times \Sigma_g}(a, \mathfrak{m})}{\del |a|}
\propto (g-1)\sum_{\alpha\in \Delta} |\alpha\cdot a|
+\sum_I \sum_{\rho_I\in \mathfrak{R}_I} 
\left( \rho_I \cdot\mathfrak{m} +(g-1)(R_I -1 ) \right) |\rho_I \cdot a |~,
\ee
where we have used the $U(1)_R$-gauge-gauge anomaly cancellation condition and
the gauge-gauge-gauge anomaly cancellation condition
\be
\sum_I \sum_{\rho_I \in \mathfrak{R}_I} \rho_I^i \rho_I^j \rho_I^k = 0~.
\ee
Thus local minimality of $V^{\rm eff}_{S^1 \times \Sigma_g}(a=0, \mathfrak{m})= 0$ is determined 
by the sign of
\be
\tilde{L}_{S^1 \times\Sigma_g}(a, \mathfrak{m})
= -\half(1-g)\sum_{\alpha\in \Delta} |\alpha\cdot a|
-\half \sum_I \sum_{\rho_I \in \mathfrak{R}_I} 
\left( -\rho_I \cdot \mathfrak{m} +(1-g)(R_I -1 ) \right) |\rho_I \cdot a |~.\label{eq:LtildeSigma}
\ee
Namely, if $\tilde{L}_{S^1 \times\Sigma_g}(a,\mathfrak{m})>0$ ($<0$),
then $V^{\rm eff}(a=0)=0$ is a local minimum (maximum), or more precisely a minimum (maximum)
in the region of $|\rho_I \cdot a|<1$ and $|\alpha\cdot a| < 1$.

%%%%%%%%%%%%%%%%%%%%%%%%%%%%%%%%%%%%%%%%%%%%%%%%%%
\subsubsection{Behavior at infinity of the $Z_{S^1 \times \Sigma_g}$ integrand}
%%%%%%%%%%%%%%%%%%%%%%%%%%%%%%%%%%%%%%%%%%%%%%%%%%
The saddle point configuration in this case is similar to the one for $T^2 \times \Sigma_g$.
Along the original integral contour of the path integral
the saddle point is given by flat connections for the gauge fields, the Coulomb branch parameters and zero modes. There are also complex saddle points with flux along $\Sigma_g$,
which can contribute to the path integral.
The localization formula of the partition function is given by \cite{Benini:2016hjo, Closset:2016arn}
\be
Z_{S^1 \times \Sigma_g}
=\sum_{\{\mathfrak{m}^i \}} \frac{1}{|W|} \oint_C d^r u\,
Z_{{\rm 1loop}\,\{\mathfrak{m}^i\}}^{S^1 \times \Sigma_g}
\Biggl[ {\rm det}_{ij}\left( 
\frac{\del^2}{\del (iu^i)\del \mathfrak{m}^j} \log{Z_{{\rm 1loop}\,\{\mathfrak{m}^i\}}^{S^1 \times \Sigma_g}}
 \right) \Biggr]^g~,
\label{eq:3dtwist}
\ee
where $u = \beta \sigma - i \beta A_Y$ and 
\begin{align}
Z_{{\rm 1loop}\,\{\mathfrak{m}^i\}}^{S^1 \times \Sigma_g} (\tau ,u)
&= Z_{{\rm vec}\,\{\mathfrak{m}^i\}}^{S^1 \times \Sigma_g} (\tau ,u ) 
Z_{{\rm chi}\,\{\mathfrak{m}^i\}}^{S^1 \times \Sigma_g} (\tau ,u )~,\\
Z_{{\rm vec}\,\{\mathfrak{m}^i\}}^{S^1 \times \Sigma_g} (\tau ,u )
&= \prod_{\alpha\in \Delta} (1-e^{i\alpha \cdot u} )^{1-g}~,\\
Z_{{\rm chi}\,\{\mathfrak{m}^i\}}^{S^1 \times \Sigma_g} (\tau ,u )
&= \prod_I \prod_{\rho_I \in \mathfrak{R}_I} 
\left( \frac{e^{i \rho_I \cdot u /2}}{1-e^{i\rho_I \cdot u}} \right)^{\rho_I \cdot \mathfrak{m}+(g-1)(R_I -1)}~.
\end{align}

For $|\sigma |\gg 1$, 
the one-loop determinants become
\begin{align}
Z_{{\rm vec}\,\{\mathfrak{m}^i\}}^{S^1 \times \Sigma_g} (\tau ,u )
 & \underset{|\sigma| \to \infty}{\longrightarrow} \prod_{\alpha\in \Delta} e^{\frac{(1-g)}{2}L|\alpha \cdot\sigma |}~,\\
Z_{{\rm chi}\,\{\mathfrak{m}^i\}}^{S^1 \times \Sigma_g} (\tau ,u ) 
& \underset{|\sigma| \to \infty}{\longrightarrow} \prod_I \prod_{\rho_I\in \mathfrak{R}_I} 
e^{\frac{- \rho_I \cdot \mathfrak{m} +(1-g)(R_I -1)}{2}L|\rho_I \cdot\sigma|}~.
\end{align}
Since the ${\rm det}_{ij}$ factor does not affect to exponential contributions,
the integrand in \eqref{eq:3dtwist} for $|\sigma |\gg 1$ behaves as
\be
Z_{{\rm 1loop}\,\{\mathfrak{m}^i\}}^{S^1 \times \Sigma_g}\underset{|\sigma| \to \infty}{\longrightarrow} \exp{\left[
- L\, \tilde{L}_{S^1 \times \Sigma_g}(\sigma, \mathfrak{m}) \right]}~,
\ee
where $\tilde{L}_{S^1 \times \Sigma_g}(a, \mathfrak{m})$ is given in \eqref{eq:LtildeSigma}.  
The condition for convergence depends on the label $\mathfrak{m}$ of the topological sector, like the effective potential. 
As in the examples above, 
the sign of $\tilde{L}_{S^1 \times \Sigma_g}$ determines both the convergence and the minimality condition for $V^{\rm eff}_{S^1\times \Sigma_g}$. 
The term $\rho_I \cdot\mathfrak{m}$ corresponds to effective CS level, while the other terms correspond to effective FI parameter.  

Note that the exponential growth of the one-loop determinant for $S^1 \times \Sigma_g$ does not imply a divergence of the localization formula itself after the integration. This is because the integral contour is taken to pick up some residues, and the exponential divergence corresponds to poles at infinity, whose residues are finite. It is interesting to see that this can shift the $\beta\rightarrow 0$ behavior of the 4d partition function.

%%%%%%%%%%%%%%%%%%%%%%%%%%%%%%%%%%%%%%%%%%%%%%%%%%
%%%%%%%%%%%%%%%%%%%%%%%%%%%%%%%%%%%%%%%%%%%%%%%%%%
\subsection{General $\mathcal{M}_3$}
%%%%%%%%%%%%%%%%%%%%%%%%%%%%%%%%%%%%%%%%%%%%%%%%%%
%%%%%%%%%%%%%%%%%%%%%%%%%%%%%%%%%%%%%%%%%%%%%%%%%%
We can obtain the condition for local minimality of $V^{\rm eff}_{\mathcal{M}_3}(a=0)=0$ for general $\mathcal{M}_3$ in a similar way.
The real part of the effective potential for generic $\mathcal{M}_3$ is given by
\begin{\eqa}
{\rm Re} (V_{\mathcal{M}_3}^{\rm eff}(a))
=-\frac{\pi^2}{2\beta} 
\sum_f (R_f L_{\mathcal{M}_3} +\rho_f \cdot l_{\mathcal{M}_3} ) 
\vartheta (\rho_f \cdot a)~.
\end{\eqa}
To see whether ${\rm Re} (V_{\mathcal{M}_3}^{\rm eff}(a=0))=0$ is a local minimum or maximum, 
we again compute the quantity
\begin{\eq}
\lim_{|a|\rightarrow +0} \frac{\del {\rm Re} (V_{\mathcal{M}_3}^{\rm eff}(a))
}{\del |a|}~,
\end{\eq}
as in the cases $\mathcal{M}_3 =L(n,1)$ and $S^1 \times \Sigma_g$~.

In the region of $|\alpha\cdot x|<1$ and $|\rho_I \cdot x|<1$ for every $\rho_I$,
we can easily calculate this as
\begin{\eq}
\frac{\del {\rm Re} (V_{\mathcal{M}_3}^{\rm eff}(a))}{\del |a|} 
=-\frac{\pi^2}{2\beta |a|} 
\sum_f (R_f L_{\mathcal{M}_3} +\rho_f \cdot l_{\mathcal{M}_3} ) |\rho_f \cdot a|~,
\end{\eq}
where we have used 
the $U(1)_R$-gauge-gauge anomaly and gauge-gauge-gauge anomaly cancellation conditions.

This means that 
the local minimality of ${\rm Re}(V_{\rm eff}(a=0))=0$ is determined by the following quantity
\begin{\eq}
\tilde{L}_{\mathcal{M}_3} (x)
= -\sum_f (R_f L_{\mathcal{M}_3} +\rho_f \cdot l_{\mathcal{M}_3} ) |\rho_f \cdot x|~.
\end{\eq}
Namely, 
if $\tilde{L}_{\mathcal{M}_3} (x)>0(<0)$ for all the directions of $x$, 
then $V_{\mathcal{M}_3}^{\rm eff} (x=0)=0$ is a local minimum (maximum).
For the $\mathcal{M}_3 =L(n,1), S^1 \times \Sigma_g$ cases,
we have seen that
the sign of $\tilde{L}_{\mathcal{M}_3} (x)>0$
determines the behavior at infinity of the integrand in the underlying 3d partition functions. Hence it is natural to expect that the sign of $\tilde{L}_{\mathcal{M}_3} (x)$ plays the same role for general $\mathcal{M}_3$. 

%%%%%%%%%%%%%%%%%%%%%%%%%%%%%%%%%%%%%%%%%%%%%%%%%%
%%%%%%%%%%%%%%%%%%%%%%%%%%%%%%%%%%%%%%%%%%%%%%%%%%
%%%%%%%%%%%%%%%%%%%%%%%%%%%%%%%%%%%%%%%%%%%%%%%%%%
\section{Minima of $V^{\rm eff}_{\CM_3}$ and Sign of ${\rm Tr}(R)$}
\label{sec:sign}
%%%%%%%%%%%%%%%%%%%%%%%%%%%%%%%%%%%%%%%%%%%%%%%%%%%%%%%%%%%%%%%%%%%%%%%%%%%%%%%%%%%%%%%%%%%%%%%%%%%%
In this section we consider several examples of asymptotically free gauge theories with charge conjugation invariance and we provide evidence for a relation between the minima of $V^{\rm eff}_{\CM_3}$ and the sign of  the anomaly ${\rm Tr}(R)$ for the non-anomalous $R$-symmetry. In all the examples the following statement holds true: $V^{\rm eff}_{\CM_3}$ has a local minimum in the origin if ${\rm Tr}(R) \leq 0$ (see the comment in footnote \ref{foot:cexample}). If more then one $R$-symmetry is preserved, the statement applies to the one that maximizes $a$. In the main text we restrict to theories with $SU(N)$ gauge group, additional examples with $SO(N)$ and $USp(2N)$ groups are considered in the appendix \ref{app:SOUSp}. Recall that, assuming there is no emergent $U(1)$ symmetry, the trace anomaly is related to the Weyl anomalies of the IR CFT by $\Tr(R) = -16(c-a)$. 

%%%%%%%%%%%%%%
%%%%%%%%%%%%%%
\subsection{Theories with Matter in a Single Representation}
%%%%%%%%%%%%%%
%%%%%%%%%%%%%%
First let us consider a 4d $\mathcal{N}=1$ theory with gauge group $G$
and only pairs of conjugate representations $(\mathfrak{R},\bar{\mathfrak{R}})$ 
(or real representations).
Since this theory has charge conjugation symmetry,
the effective potential is 
\begin{\eq}
V_{\mathcal{M}_3}^{\rm eff}(a )
=-\frac{\pi^2 L_{\mathcal{M}_3}}{2\beta} 
\Biggl[ \sum_{\alpha}\vartheta (\alpha\cdot a )
 +2\sum_{I\in\mathfrak{R}} (R_I -1) \sum_{\rho\in\mathfrak{R}} \vartheta (\rho\cdot a )
 \Biggr]~.
\end{\eq}
Because of the $U(1)_R \times G\times G$ anomaly cancellation condition
\begin{\eq}
T({\rm adj}) 
+2 T(\mathfrak{R}) \sum_{I\in \mathfrak{\mathfrak{R}}} (R_I -1) =0~,
\end{\eq}
where $T(\mathfrak{R})$ is the quadratic Dynkin index,
${\rm Tr}(R)$ of this theory is uniquely determined as
\begin{\eq}
{\rm Tr}(R)
= {\rm dim(adj)} 
-\frac{T({\rm adj})}{T(\mathfrak{R})} {\rm dim}(\mathfrak{R})~.\label{eq:onlyRtrace}
\end{\eq}
The anomaly cancellation simplifies also 
the effective potential and $\tilde{L}_{\mathcal{M}_3}$ as\footnote{
Note that these expressions are true also for
theories with only real representations $\mathfrak{R}$.
}
\begin{\eqa}
&& V_{\mathcal{M}_3}^{\rm eff}(a )
=\frac{\pi^2 L_{\mathcal{M}_3}}{2\beta} \Biggl[
 -\sum_{\alpha}\vartheta (\alpha\cdot a ) 
 +\frac{T({\rm adj})}{T(\mathfrak{R})} 
 \sum_{\rho\in\mathfrak{R}} \vartheta (\rho \cdot a )  
\Biggr]~,\NN\\
&& \tilde{L}_{\mathcal{M}_3}(a)
=-\sum_{\alpha} |\alpha\cdot a | 
  +\frac{T({\rm adj})}{T(\mathfrak{R})}
  \sum_{\rho\in\mathfrak{R}} |\rho \cdot a |~. \label{eq:onlyRV}
\end{\eqa}

%%%%%%%%%%%%%%
\subsubsection{Theories with adjoint representations}
\label{sec:adj}
%%%%%%%%%%%%%%
For this case,
the $U(1)_R$-gauge-gauge anomaly cancellation forces us to
\begin{\eq}
{\rm Tr}(R)=0~,\quad 
V^{\rm eff}_{\mathcal{M}_3}(a)=0~.
\end{\eq}
Hence
the minimum of the effective potential is trivially zero 
but not isolated.
Namely
the effective potential has flat directions and
the $\beta\rightarrow 0$ limit of $Z_{S^1 \times\mathcal{M}_3}$
receives $\mathcal{O}(\beta^r )$ correction.

The flatness of the potential is consistent with the fact that center symmetry remains unbroken for gauge theories with adjoint matter (even without supersymmetry) and periodic conditions for the fermions \cite{Kovtun:2007py, Unsal:2007vu}.

%%%%%%%%%%%%%%
\subsubsection{$SU(N)$ theory with fundamentals}
%%%%%%%%%%%%%%
Noting $T({\rm adj})=2N$ and $T({\rm fund})=1$,
we find
\begin{\eqa}
&& {\rm Tr}(R) =-N^2 -1 <0~,\NN\\
&& V_{\mathcal{M}_3}^{\rm eff}(a )
=\frac{\pi^2 L_{\mathcal{M}_3}}{2\beta} \Biggl[ -\sum_{1\leq i\neq j \leq N}\vartheta (a_i -a_j  ) 
 +2N \sum_{i=1}^N \vartheta ( a_i )   \Biggr]~,
\end{\eqa}
with the constraint $\sum_{i=1}^N a_i =0$.
In order to find the minimum of the potential,
we use the following inequality \cite{2006math......7093R}
as in \cite{Ardehali:2015bla}
\begin{\eq}
\sum_{1\leq i,j\leq n}\vartheta (c_i -d_j )
-\sum_{1\leq i<j\leq n}\vartheta (c_i -c_j )
-\sum_{1\leq i<j\leq n}\vartheta (d_i -d_j )
\geq \vartheta \left(  \sum_{i=1}^n (c_i -d_i ) \right)~.
\label{ineq:theta}
\end{\eq}
Taking $c_i =a_i$ and $d_i =0$ in this inequality,
we find
\begin{\eq}
2N\sum_{i=1}^N \vartheta ( a_i  )
-\sum_{1\leq i\neq j\leq n}\vartheta (a_i -a_j )
\geq 0~,
\label{ineq:f}
\end{\eq}
which leads us to
\begin{\eq}
V_{\mathcal{M}_3}^{\rm eff}(a ) \geq 0~.
\end{\eq}
Since $V_{\mathcal{M}_3}^{\rm eff}(a )=0$ is realized by $a_i =0$,
the effective potential has a global minimum in the origin. 

%%%%%%%%%%%%%%
%%%%%%%%%%%%%%
\subsubsection{$SU(N)$ theory with two-index symmetric representation $S_2$}
\label{sec:SUS2}
%%%%%%%%%%%%%%
%%%%%%%%%%%%%%
Let us consider $SU(N)$ theories with pairs of two-index symmetric representation $S_2$ and its conjugate.
Since $T(S_2 )=N+2$,
${\rm Tr}(R)$ of this theory is given by
\begin{\eq}
{\rm Tr}(R) = \frac{(N+1)(N-2)}{N+2} \geq 0~.
\end{\eq}
Note that 
${\rm Tr}(R)$ vanishes for $N=2$ 
while ${\rm Tr}R>0$ for higher $N$ 
since $S_2$ is the adjoint representation for the $SU(2)$ case.
Since the $SU(2)$ case is the special case of sec.~\ref{sec:adj},
below we take $N\geq 3$.
The effective potential after the anomaly cancellation is
\begin{\eqa}
V_{\mathcal{M}_3}^{\rm eff}(a )
=\frac{\pi^2 L_{\mathcal{M}_3}}{2\beta} \Biggl[
 -\sum_{1\leq i\neq j\leq N}\vartheta (a_i -a_j  ) 
 +\frac{2N}{N+2} \sum_{1\leq i\leq j\leq N} \vartheta ( a_i +a_j )  \Biggr]~,
\end{\eqa}
with the constraint $\sum_{i=1}^N a_i =0$.
We can easily see that 
this potential has a nontrivial minimum for $N\geq 3$.
To see this, it is convenient to look at $\tilde{L}_{\mathcal{M}_3}$:
\begin{\eqa}
\tilde{L}_{\mathcal{M}_3}(a )
= -\sum_{1\leq i\neq j\leq N} |a_i -a_j | 
 +\frac{2N}{N+2} \sum_{1\leq i\leq j\leq N} |a_i +a_j |~.
\end{\eqa}
As we have shown in sec.~\ref{sec:min},
the potential does not have a local minimum in the origin
if $\tilde{L}_{\mathcal{M}_3}(a )$ is negative along at least one direction.
For example,
if we take $a_1 =a =-a_N$ and $a_{j=2,\cdots ,N-1}={\rm finite}$ with $a\gg 1$,
then we have
\begin{\eq}
\left. \tilde{L}_{\mathcal{M}_3}(a) \right|_{a_1 =a= -a_N, a\gg 1}
= -\frac{4(N-2)}{N+2} a <0~.
\end{\eq}
Therefore $V_{\mathcal{M}_3}^{\rm eff}(a=0 )=0$ is not a local minimum and
the potential has a nontrivial minimum outside the origin.

%%%%%%%%%%%%%%
%%%%%%%%%%%%%%
\subsubsection{$SU(N)$ theory with three-index symmetric representation $S_3$}
%%%%%%%%%%%%%%
%%%%%%%%%%%%%%
The $SU(N)$ theory with $N_f$ pairs of three index symmetric representation $S_3$ and its conjugate
has non-positive beta function when $N_f \leq 6N/(N+2)(N+3)$.
Hence we cannot have nonzero $N_f$ for $N\geq 3$ and
the only possible choice is the $SU(2)$ theory with one three index symmetric representation,
which is called ISS model \cite{Intriligator:1994rx}.
Since $T(S_3 )=(N+3)(N+2)/2$,
${\rm Tr}(R)$ and the effective potential of the ISS model are given by
\begin{\eq}
 {\rm Tr}(R) 
= \frac{7}{5} >0~,\quad
V_{\mathcal{M}_3}^{\rm eff}(a )
=\frac{\pi^2 L_{\mathcal{M}_3}}{\beta} \Biggl[
-\vartheta (2a  ) 
 +\frac{2}{5}\left( \vartheta (a) +\vartheta (3a) \right) 
 \Biggr]~.
\end{\eq}
As in the case $\mathcal{M}_3 =S^3$ \cite{Ardehali:2015bla},
the minimum of the effective potential is given by $a=\pm 1/3$ and 
the partition function in the $\beta\rightarrow 0$ limit is
\be
\log{Z_{S^1 \times \CM_3}} \underset{\beta \to 0}{\longrightarrow} -\frac{\pi^2 L_{\CM_3}}{12 \beta}\left( \Tr(R) -\frac{8}{5} \right)
 +\,\dots~.
\ee

%%%%%%%%%%%%%%

%%%%%%%%%%%%%%
%%%%%%%%%%%%%%
\subsection{Theories with Additional Adjoint Matter}
%%%%%%%%%%%%%%
%%%%%%%%%%%%%%
Next we consider a theory
with gauge group $G$, adjoint representations and
pairs of conjugate representations $(\mathfrak{R},\bar{\mathfrak{R}})$ (or real representations).
The effective potential of this theory is 
\begin{\eqa}
V_{\mathcal{M}_3}^{\rm eff}(a )
&=&\frac{\pi^2 L_{\mathcal{M}_3}}{2\beta} \Biggl[ 
 -\sum_{\alpha}\vartheta (\alpha\cdot a )
 -\sum_{I\in {\rm adj}} (R_I -1)
    \sum_{\alpha}\vartheta (\alpha\cdot a ) \NN\\
&& -2\sum_{I\in \mathfrak{R}}(R_I -1)\sum_{\rho\in\mathfrak{R}} \vartheta (\rho\cdot a ) \Biggr]~.
\end{\eqa}
By the $U(1)_R \times G\times G$ anomaly cancellation condition
\begin{\eq}
T({\rm adj})\left( 1 
+\sum_{I\in {\rm adj} } (R_I -1) \right)
+2T(\mathfrak{R})  \sum_{I\in \mathfrak{R}} (R_I -1) =0~,
\end{\eq}
${\rm Tr}(R)$ and the effective potential are simplified as
\begin{\eqa}
&&{\rm Tr}(R)
= \left({\rm dim(adj)}   -  \frac{T({\rm adj})}{T(\mathfrak{R})}{\rm dim}(\mathfrak{R}) \right) \left( 1 
+\sum_{I\in {\rm adj} } (R_I -1) \right) , \\
&& V_{\mathcal{M}_3}^{\rm eff}(a )
= \frac{\pi^2 L_{\mathcal{M}_3} {\rm Tr}(R)}
{ 2\beta ({\rm dim(adj)} 
-\frac{T({\rm adj})}{T(\mathfrak{R})} {\rm dim}(\mathfrak{R}) )} 
\Biggl[ - \sum_{\alpha}\vartheta (\alpha\cdot a ) 
 + \frac{T({\rm adj})}{T(\mathfrak{R})} \sum_{\rho\in \mathfrak{R}} \vartheta (\rho \cdot a )  \Biggr]~. \NN
\end{\eqa}
Similarly, $\tilde{L}_{\mathcal{M}_3}(a)$ becomes
\begin{\eq}
\tilde{L}_{\mathcal{M}_3}(a)
=\frac{{\rm Tr}(R)}{ {\rm dim(adj)} 
-\frac{ T({\rm adj})}{T(\mathfrak{R})} {\rm dim}(\mathfrak{R})} 
\Biggl[- \sum_{\alpha} |\alpha\cdot a | 
  + \frac{T({\rm adj})}{T(\mathfrak{R})} \sum_{\rho\in\mathfrak{R}} |\rho \cdot a |  \Biggr] .
\end{\eq}
Comparing with eq.s \eqref{eq:onlyRtrace}-\eqref{eq:onlyRV} we see that
\be
\left(V^{\rm eff}_{\CM_3}/\Tr(R)\right)_{(\mathfrak{R},\bar{\mathfrak{R}})\oplus {\rm adj}} = \left(V^{\rm eff}_{\CM_3}/\Tr(R)\right)_{(\mathfrak{R},\bar{\mathfrak{R}})}~,\label{eq:adjmatt}
\ee
and similarly for $\tilde{L}_{\CM_3}$. 

Consider the case in which the theory before the addition of adjoint matter has $\Tr(R)$ negative, and a local minimum in zero. Then, using the relation \eqref{eq:adjmatt} we see that the theory with adjoint matter will either have $\Tr(R)$ still negative, and still have a local minimum of the potential in the origin, or $\Tr(R)$ will flip sign, and then the potential will have a local maximum in the origin. 

It remains to consider the case in which the theory without adjoint matter has $\Tr(R)$ positive. If we add adjoint matter and the sign of $\Tr(R)$ becomes negative, we cannot prove the existence of a local minimum in the origin. In all the examples in the previous section and in the appendix \ref{app:SOUSp} that have $\Tr(R) > 0$, we have checked that upon addition of adjoint matter, $a$-maximization gives still $\Tr(R) > 0$.

We conclude that in all the examples that we consider, we can add adjoint matter and  the relation between the minimum of the potential and the sign of $\Tr(R)$ still holds.

%%%%%%%%%%%%%%%%%%%%%%%%%
\section{Discussion}\label{sec:disc}
%%%%%%%%%%%%%%%%%%%%%%%%%%%

In this work we considered the $\beta \to 0$ limit of the localization integrand for $4d$ $\CN = 1$ theory on $S^1 \times \CM_3$, providing a new derivation of the resulting effective potential for the holonomies around $S^1$ and generalizing the results of \cite{Ardehali:2015bla}. We have seen that theories fall into two qualitatively different classes depending on whether or not the effective potential is minimized in the origin, where it vanishes. If it is, the leading Cardy-like behavior of the $4d$ partition function is fixed by the $\Tr(R)$ anomaly. On the other hand, cases with a nontrivial negative minimum outside of the origin are such that the localization integrand for the dimensionally-reduced $3d$ theory grows exponentially at infinity, and their Cardy-like behavior is not determined just by $\Tr(R)$. We checked in several examples that the sign of $\Tr(R)$ is negative for theories with a minimum in the origin.

There are some questions about the effective potential for the holonomies that are left unanswered. Firstly, it would be interesting to prove the connection between the minimum and the sign of $\Tr(R)$, or find a counterexample. In the examples considered in this work, we have only checked that when $\Tr(R) > 0$ the origin is not a minimum. One thing to do is to try and find the nontrivial minimum in these cases, and see if the correction to the Cardy-like behavior is such that $Z_{S^1 \times \CM_3} \underset{\beta \to 0}{\longrightarrow} \infty$ (recall that in this case, since $\Tr(R) > 0\,$, \eqref{eq:Cardy} without corrections would give $Z_{S^1 \times \CM_3} \underset{\beta \to 0}{\longrightarrow} 0$). Moreover, for the cases without charge conjugation symmetry $\rho_f \leftrightarrow - \rho_f$, the analysis of the asymptotic limit of the integral over holonomies is complicated by the additional terms proportional to the densities $A_{\CM_3}$ and $l^i_{\CM_3}$, and it has not been studied here. The combination $c-a$ plays an important role in holographic theories \cite{Camanho:2014apa}, and it would be worth exploring if there is a connection. Formulas for $c-a$ in terms of short multiplets in the bulk have appeared in \cite{Ardehali:2013gra, Ardehali:2013xya, Beccaria:2014xda}.

Finally, it would be interesting to explore the interpretation of the correction $V^{\rm eff}_{\CM_3}(a_{\rm min})$ to the Cardy-like behavior, from the point of view of the IR fixed point. Related to this, one should bear in mind that the exponential growth of the integrand in the $3d$ partition function is not an intrinsic property of the fixed point, but rather it depends on the particular weakly-coupled UV completion used to write down the localization formula. For instance, in the $3d$ SQCD-like theories considered in \cite{Safdi:2012re, Lee:2016zud}, the dual weakly-coupled UV completions give localization formulas with different domains of convergence as a function of the choice of $R$-symmetry. Given the analytic properties of $Z_{\CM_3}$ \cite{Closset:2014uda}, even when the matrix model is divergent one should be able to assign a finite value to the partition function via analytic continuation. One can imagine a situation in which a $4d$ theory with $V^{\rm eff}_{\CM_3}(a_{\rm min}) < 0$ and divergent dimensionally-reduced partition function admits a dual weakly-coupled description with $V^{\rm eff}_{\CM_3}(a_{\rm min}= 0) = 0$ and finite dimensionally-reduced partition function. It is amusing to observe that in this situation there would be a mismatch in the leading behavior of the index for the two dual descriptions, which should be ascribed to the existence of an emergent symmetry that mixes with the $R$-symmetry. In this hypothetical situation, $V^{\rm eff}_{\CM_3}(a_{\rm min})$ would be interpreted as the trace anomaly of the $U(1)$ that mixes with the $R$-symmetry. Unfortunately, in the known examples of theories with a nontrivial minimum we do not know a dual weakly-coupled description.

We hope to come back to these questions in the future.

%%%%%%%%%%%%%%%%%%%%%%%%%%%%%%%%%%%%%%%%%%%%%%%%%%%
\subsection*{Acknowledgments}
%%%%%%%%%%%%%%%%%%%%%%%%%%%%%%%%%%%%%%%%%%%%%%%%%%
We thank Jaume Gomis, Heeyeon Kim, Zohar Komargodski, Itamar Shamir and Yutaka Yoshida for useful discussions and comments. Research
at Perimeter Institute is supported by the Government of Canada through Industry
Canada and by the Province of Ontario through the Ministry of Research \& Innovation.

%%%%%%%%%%%%%%%%%%%%%%%
%
%
\appendix

\section{Supersymmetric Chern-Simons actions}
\label{app:CSactions}
The gauge-gauge, gauge-$R$ and gauge-KK supersymmetric Chern-Simons actions can be found in \cite{Closset:2012vp}. The $R$-KK and KK-KK supersymmetric Chern-Simons terms can be obtained from \cite{Kuzenko:2013uya}.\footnote{There are minor differences between the conventions of \cite{Closset:2012vp} and ours, namely $V^{\text{There}}_\mu = v^{\text{Here}}_\mu$, $A_\mu^{\text{There}} = \CA_\mu^{(R)\text{Here}}$, and they define the Ricci scalar to be negative for a sphere, while here we take it to be positive, therefore $R^{\text{There}} = - R^{\text{Here}}$. The relation between the conventions in \cite{Kuzenko:2013uya} and ours is:  $Z^{\text{There}} = 2 H^{\text{Here}}$, $H_\mu^{\text{There}} = - 2 v_\mu^{\text{Here}}$, $b_\mu^{\text{There}} = (\CA^{(R)}_\mu - \frac 32 v_\mu)^{\text{Here}}$ and finally $a_\mu^{\text{There}} = 2 c_\mu^{\text{Here}}$.}

We list here the supersymmetric Chern-Simons actions and we substitute the configuration \eqref{eq:susy3dvec} for the vector multiplet (the substitution is denoted with an arrow):
\begin{itemize}
\item{Gauge-gauge 
\begin{align}
& \int_{\CM_3}d^3 x \sqrt{h}  \left[i \epsilon_{\mu\nu\rho}(\rho_f\cdot \CA)^\mu \partial^\nu (\rho_f \cdot \CA)^\rho  - 2 (\rho_f\cdot D)(\rho_f \cdot \sigma) \right]\nonumber\\  \longrightarrow~~ &\frac{4 \pi^2}{\beta^2}(\rho_f\cdot a)^2\int_{\CM_3}d^3 x \sqrt{h}  \left[-c^\mu v_\mu + 2 H \right] \nonumber \\ - & \frac{2\pi}{\beta} 2 (\rho_f\cdot a)\int_{\CM_3}d^3 x \sqrt{h} \left[-(\rho_f \cdot A^\mu)v_\mu+  (\rho_f \cdot D)\right] + \mathcal{O}(\beta^0)~.\label{eq:AA}
\end{align}
}
\item{Gauge-$R$ 
\begin{align}
&\int_{\CM_3}d^3 x \sqrt{h}  
\left[i \epsilon_{\mu\nu\rho}(\rho_f\cdot \CA)^\mu \partial^\nu (\CA^{(R) \rho} - \tfrac 12 v^\rho)  - (\rho_f\cdot D) H 
-\frac{(\rho_f \cdot \sigma)}{4} 
(R  + 2 v_\mu v^\mu + 2 H^2) \right]\nonumber\\  
\longrightarrow - & \frac{2 \pi}{\beta} (\rho_f \cdot a)\int_{\CM_3}d^3 x \sqrt{h}  \left[-\CA^{(R)\mu} v_\mu  +  v^\mu v_\mu - \half H^2  + \frac 14 R \right] + \mathcal{O}(\beta^0)~.\label{eq:AR}
\end{align}
}
\item{Gauge-KK 
\begin{align}
&\frac{2\pi}{\beta}\int_{\CM_3}d^3 x \sqrt{h}  \left[i \epsilon_{\mu\nu\rho}(\rho_f\cdot \CA)^\mu \partial^\nu c^\rho  + (\rho_f\cdot D)  -(\rho_f \cdot \sigma) H \right]\nonumber\\  \longrightarrow - &\frac{4 \pi^2}{\beta^2}(\rho_f \cdot a)\int_{\CM_3}d^3 x \sqrt{h}  \left[-c^\mu v_\mu + 2 H \right] \nonumber \\+ & \frac{2 \pi}{\beta}\int_{\CM_3}d^3 x \sqrt{h} \left[-(\rho_f \cdot A^\mu)v_\mu +  (\rho_f \cdot D)\right] + \mathcal{O}(\beta^0)~.\label{eq:Ac}
\end{align}
}
\item{$R$-KK
\begin{align}
& \frac{2\pi}{\beta}\int_{\CM_3}d^3 x \sqrt{h}  \left[-\CA^{(R)\mu} v_\mu  +  v^\mu v_\mu - \half H^2  + \frac 14 R \right]~.\label{eq:Rc}
\end{align}
}
\item{KK-KK
\begin{align}
&\frac{4 \pi^2}{\beta^2}\int_{\CM_3}d^3 x \sqrt{h}  \left[-c^\mu v_\mu+ 2 H \right]~.\label{eq:cc}
\end{align}
}
\end{itemize} 

\section{Special Functions}\label{app:SpF}

\subsection{Elliptic Gamma Function}
The elliptic Gamma function is defined as 
\be
\Gamma_e (z;p ,q )
= \prod_{j,k\geq 0} \frac{1 -z^{-1}p^{j+1}q^{k+1}}{1 -zp^{j}q^{k}}~.\label{eq:EllG}
\ee
We also use the notation $\Gamma(a ; \sigma, \tau) \equiv \Gamma_e(e^{2 \pi i a}; e^{2 \pi i \sigma}, e^{2\pi i \tau})$. 

Setting $\tau = b_1 \beta$, $\sigma = b_2 \beta$ we take the limit $\beta \to 0$ with $b_{1,2}$ fixed. The limit is given by the following identity \cite{2006math......7093R} 
\be
\ln \Gamma \left( x+\frac{R(\sigma +\tau )}{2};\sigma ,\tau  \right)
\underset{\beta \to 0}{\longrightarrow} 2\pi i \Biggl[ -\frac{\kappa (x)}{12\tau\sigma}
+(R -1) \frac{\tau +\sigma}{4\tau\sigma} \left( \vartheta (x) -\frac{1}{6}\right) \Biggr] +\mathcal{O}(\beta^0 )~,\label{eq:LimEllG}
\ee
where $\kappa$ and $\vartheta$ are defined in \eqref{eq:kappa} and \eqref{eq:vartheta}. To study the asymptotic on $L(n,1)$, it is convenient to rearrange the identity \eqref{eq:LimEllG} in the following form
\begin{align}
& \ln \Gamma \left( x+\frac{R (\sigma +\tau )}{2} +m\sigma ;\sigma +\tau ,n\sigma  \right) \label{eq:LimEllGn1}\\
&\underset{\beta \to 0}{\longrightarrow}  \frac{2\pi i}{n} \Biggl[ -\frac{\kappa (x)}{12\sigma (\sigma +\tau )}
+ \frac{ R (\tau+\sigma ) +(2m-n-1)\sigma -\tau }{4\sigma (\sigma +\tau )} 
\left( \vartheta (x) -\frac{1}{6}\right) \Biggr] +\mathcal{O}(\beta^0 )~,\nonumber\\
& \ln \Gamma \left( x+\frac{R (\sigma +\tau )}{2} +(n-m)\tau ;\sigma +\tau ,n\tau  \right)  \label{eq:LimEllGn2}\\
&\underset{\beta \to 0}{\longrightarrow} \frac{2\pi i}{n} \Biggl[ -\frac{\kappa (x)}{12\tau (\sigma +\tau )}
+ \frac{ R (\tau+\sigma ) +(n-2m-1)\tau -\sigma }{4\tau (\sigma +\tau )} 
\left( \vartheta (x) -\frac{1}{6}\right) \Biggr] +\mathcal{O}(\beta^0 )~.\nonumber
\end{align}

\subsection{Dedekind Eta Function and Jacobi Theta Function}
The Dedekind eta function $\eta (\tau )$ and the Jacobi theta function
$\theta_1 (\tau |z)$ are defined as follows 
\be
\eta (\tau )
= q^{1/24} \prod_{k=0}^\infty (1 -q^{k+1} ) ,\quad
\theta_1 ( \tau | z)
= -iq^{\frac{1}{8}}y^{\frac{1}{2}}
 \prod_{k=1}^\infty (1-q^k )(1-yq^k ) (1-y^{-1}q^{k-1})~, \label{eq:EtaTheta}
\ee
with $q=e^{2\pi i\tau}$ and $y = e^{2 \pi iz}$. Setting $\tau = i \beta / L$ and $z = a$ we take the limit $\beta \to 0$ with $L$ and $a$ fixed. Using the following identity
\begin{align}
& \theta_1 (\tau | z +n)=(-1)^n \theta_1 (\tau |z)~,~~{\rm for}~n\in\mathbb{Z}~, \\
&\eta (\tau ) = \frac{1}{\sqrt{-i\tau}} \eta (-\tfrac{1}{\tau} )~,\\
& \theta_1 (\tau |z) = \frac{i}{\sqrt{-i\tau}}e^{-\frac{\pi iz^2}{\tau}}
 \theta_1 (-\tfrac{1}{\tau} |\tfrac{z}{\tau})~,
\end{align}
we see that the limit is
\begin{align}
\eta (\tau ) 
& \underset{\beta \to 0}{\longrightarrow}
\sqrt{\frac{L}{\beta}}\displaystyle{ e^{-\frac{1}{24}\frac{2\pi L}{\beta} + \CO(\beta^0)}}~, \label{eq:EtaLim}\\
\theta_1 (\tau |a ) 
& \underset{\beta \to 0}{\longrightarrow}
(-1)^{[a]} \sqrt{\frac{L}{\beta}}
\displaystyle{ e^{-\frac{1}{8}\frac{2\pi L}{\beta}-\frac{\pi L}{\beta}\{  a \}^2+\frac{\pi L}{\beta}\{  a  \} + \CO (\beta^0)} }~.\label{eq:ThetaLim}
\end{align}

\subsection{Double Sine Function}
The double sine function $s_b(z)$ is defined by
\be
s_b (z)
= \prod_{m,n=0}^\infty \frac{mb+nb^{-1} +\frac{b+b^{-1}}{2} -iz}
  {mb+nb^{-1} +\frac{b+b^{-1}}{2} +iz}~.\label{eq:DouSin}
\ee
The following identity holds true
\be
- \log{s_b (x-i(1-\Delta )/2 ) }
\underset{|x| \to \infty}{\longrightarrow} - \frac{i\pi {\rm sgn}(x)}{2}x^2 
-\frac{\pi(1-\Delta)}{2}|x| +\mathcal{O}(|x|^0 )~.
\label{eq:DouSinLim}
\ee

%%%%%%%%%%%%%%%%%%%%%%
\section{More on Minima of $V^{\rm eff}$ and Sign of $\Tr(R)$}\label{app:SOUSp}
%%%%%%%%%%%%%%%%%%%%%%%
%%%%%%%%%%%%%%%%%%%%%%%%
\subsection{$SO(N)$ Theories}
%%%%%%%%%%%%%%%%%%%%%%%%
\subsubsection{$SO(N)$ theory with fundamentals}
%%%%%%%%%%%%%%%%%%%%%%%%%
\label{sec:SOfund}
%%%%%%%%%%%%%%
Let us consider $SO(N)$ theories with $N\geq 3$ and 
fundamental chiral multiplets.
Using $T({\rm adj})=2(N-2)$ and $T({\rm fund})=2$,
${\rm Tr}(R)$ of this theory is
\begin{\eqa}
{\rm Tr}(R) = -\frac{N (N-3)}{2} \leq 0~.
\end{\eqa}
The effective potential has 
the same expression both for $SO(2n)$ and $SO(2n+1)$
\begin{\eqa}
V_{\rm eff}(a )
=\frac{\pi^2 L_{\mathcal{M}_3}}{  \beta} \Biggl[
 -\sum_{1\leq i < j\leq n}(\vartheta (a_i -a_j )
  + \vartheta (a_i +a_j ))
 + 2 (n-1) \sum_{i=1}^n \vartheta ( a_i ) \Biggr]~.
\label{eq:SOf}
\end{\eqa}

It was shown for the $SO(2n+1)$ case in \cite{Ardehali:2015bla} that
the minimum of this potential is zero and
given by the configuration with only one nonzero component $a_i$.
Hence although the origin is a minimum, it is not isolated:
the potential has one-dimensional flat directions and the partition function in the $\beta\rightarrow 0$ limit has $\mathcal{O}(\beta )$ corrections.

%%%%%%%%%%%%%%
\subsubsection{$SO(N)$ theory with two-index symmetric representation $S_2$}
%%%%%%%%%%%%%%
The $SO(N)$ theory with $N_f$ two-index symmetric representations
has non-positive beta function for $N_f \leq 3(N-2)/(N+2)$.
The $N_f =1$ case is known as BCI model \cite{Brodie:1998vv},
which was analyzed in \cite{Ardehali:2015bla} 
for the $N=2n+1$ case.
Since $T(S_2)=2(N+2)$,
${\rm Tr}(R)$ in this class of theory is given by
\begin{\eq}
{\rm Tr}(R)
= N - 1 >0~.
\end{\eq}
The effective potential for $SO(2n)$ case is
\begin{\eq}
V_{\mathcal{M}_3}^{\rm eff}(a )
=\frac{2 \pi^2 L_{\mathcal{M}_3}}{\beta(n+1)}\Biggl[
  - \sum_{1\leq i < j\leq n}
 \left( \vartheta (a_i -a_j )  +\vartheta (a_i +a_j ) \right)
      +\frac{n-1}{2} \sum_{i=1}^n \vartheta (2a_i ) 
\Biggr]~,
\end{\eq}
while the one for $SO(2n+1)$ case is
\begin{\eq}
V_{\mathcal{M}_3}^{\rm eff}(a )
=\frac{2 \pi^2 L_{\mathcal{M}_3}}{\beta(n + \frac 32)} \Biggl[
  -\sum_{1\leq i < j\leq n}
 \left( \vartheta (a_i -a_j ) +\vartheta (a_i +a_j ) \right) 
 +  \sum_{i=1}^n 
 \left(-\vartheta (a_i ) + \frac{n-\half}{2}\vartheta (2a_i ) \right)
 \Biggr]~.
\end{\eq}
Both effective potentials have a nontrivial minimum,
because the corresponding $\tilde{L}$ functions are negative along $a_1 \gg 1$ direction
\begin{align}
\left. \tilde{L}_{\mathcal{M}_3}(a) \right|_{SO(2n), a_1 \gg 1}
& = -\frac{4(n-1)}{n+1}a_1 <0\,,~\\
\left. \tilde{L}_{\mathcal{M}_3}(a) \right|_{SO(2n+1), a_1 \gg 1}
 & = -\frac{4(n - \half )}{n+\frac 32 }a_1  <0~.
\end{align}
Hence the origin does not give a local minimum of the potentials.
According to the analysis in \cite{Ardehali:2015bla} for the $SO(2n+1)$ case,
the effective potential actually attains the minimum 
when $[(3n+1)/4]$ of $a_i$ are $\pm 1/2$, and the other components are zero:
\begin{\eq}
\left. V_{\mathcal{M}_3}^{\rm eff}(a_{\rm min} ) \right|_{SO(2n+1)}
= -\frac{2\pi^2 L_{\mathcal{M}_3}}{\beta(n+\frac 32)} \sum_{1\leq j\leq \left[ \frac{3n+1}{4} \right]} (3n+1-4j)~.
\end{\eq}
This contributes to the order $\mathcal{O}(\beta^{-1})$ of the logarithm of the partition function in the $\beta\rightarrow 0$ limit.

%%%%%%%%%%%%%%
\subsection{$USp(2N)$ Theories}
\subsubsection{$USp(2N)$ theory with fundamentals}
\label{sec:USp}
%%%%%%%%%%%%%%
Since  $T({\rm adj})=2(N+1)$ and $T({\rm fund})=1$,
we find
\begin{\eqa}
&& {\rm Tr}(R) = -N(2N+3) < 0~,\\
&& V_{\mathcal{M}_3}^{\rm eff} (a )
= \frac{\pi^2 L_{\mathcal{M}_3}}{\beta} \Biggl[
 -\sum_{1\leq i < j\leq N} \left( \vartheta (a_i -a_j ) +\vartheta (a_i +a_j ) \right)
 + \sum_{i=1}^N \left( - \vartheta ( 2a_i )   +2(N+1) \vartheta ( a_i ) \right) \Biggr]~. \NN
\label{eq:USp}
\end{\eqa}
From the result for the $SO(N)$ theories with fundamentals in sec.~\ref{sec:SOfund},
we have the following inequality
\begin{\eq}
-\sum_{1\leq i < j\leq N} \left( \vartheta (a_i -a_j )  +\vartheta (a_i +a_j ) \right)
 +2 (N-1) \sum_{i=1}^N \vartheta ( a_i ) \geq 0~.
\end{\eq}
Plugging this into the effective potential leads us to
\begin{\eqa}
V_{\mathcal{M}_3}^{\rm eff}(a )
\geq   \frac{\pi^2 L_{\mathcal{M}_3}}{\beta}\sum_{i=1}^N \left( -  \vartheta ( 2a_i )  +4 \vartheta ( a_i ) \right) \geq 0~.
\end{\eqa}
Thus the minimum of the effective potential is zero and realized by the origin $a_i =0$.

 %%%%%%%%%%%%%%
\subsubsection{$USp(2N)$ theory with two-index anti-symmetric representation $A_2$}
%%%%%%%%%%%%%%
Let us take $N\geq 2$.
Using  $T(A_2 )=2(N-1)$,
${\rm Tr}(R)$ and the potential are
\begin{\eqa}
&& {\rm Tr}(R) = -N< 0~,\\
&& V_{\mathcal{M}_3}^{\rm eff} (a )
= \frac{\pi^2 L_{\mathcal{M}_3}}{2\beta} \Biggl[
 -\frac{2N-3}{N-1} \sum_{1\leq i\neq j\leq N} \left( \vartheta (a_i -a_j ) +\vartheta (a_i +a_j ) \right)
 -2 \sum_{i=1}^N  \vartheta ( 2a_i )   \Biggr]~. \NN
\end{\eqa}
Applying the inequality (C.10), we obtain
\begin{\eqa}
V_{\mathcal{M}_3}^{\rm eff}(a )
\geq   \frac{\pi^2 L_{\mathcal{M}_3}}{\beta}\sum_{i=1}^N \left( 2(2N-3) \vartheta ( a_i )   -\vartheta ( 2a_i )  \right) \geq 0~.
\end{\eqa}
Therefore the minimum of the effective potential is zero.

%BIBLIOGRAPY
\bibliographystyle{JHEP}
\bibliography{draft}
%\begin{thebibliography}{99}

\end{document}